\documentclass[usenatbib]{mn2e}

\newcommand{\etal}{{et al.}}
\newcommand{\ie}{{\it i.e.}}

\newcommand{\be}{\begin{equation}}
\newcommand{\ee}{\end{equation}}

\newcommand{\kmsMpc}{km~s$^{-1}$ Mpc$^{-1}$}

\newcommand{\mnras}{MNRAS}
\newcommand{\apj}{ApJ}
\newcommand{\apjl}{ApJL}
\newcommand{\aj}{AJ}

\newcommand{\apjs}{ApJS}

\newcommand{\nmatch}{240~}

\usepackage{color}
\definecolor{red}{rgb}{0.7,0.1,0.1}
\definecolor{blue}{rgb}{0.1,0.1,0.7}
\definecolor{black}{rgb}{0.0,0.0,0.0}
\definecolor{green}{rgb}{0.1,0.7,0.1}
\newcommand{\klm}{\textcolor{black}}

\usepackage{epsfig}
\begin{document}

\title[The Morphology of BOSS Galaxies]{The Morphology of Galaxies in the Baryon Oscillation Spectroscopic Survey}
\author[K.L. Masters \etal]{Karen L. Masters$^{1,2}$, Claudia Maraston$^1$, Robert C. Nichol$^{1,2}$, Daniel Thomas$^{1,2}$, \newauthor Alessandra Beifiori$^1$, Kevin Bundy$^3$, Edward M. Edmondson$^1$, Tim D. Higgs$^1$,  \newauthor Alexie Leauthaud$^{4,5}$,   Rachel Mandelbaum$^6$, 
Janine Pforr$^1$, Ashley J. Ross$^1$,  \newauthor Nicholas P. Ross$^4$, Donald P. Schneider$^7$, Ramin Skibba$^8$,  Jeremy Tinker$^9$, \newauthor  Rita Tojeiro$^1$, David A. Wake$^{10}$, Jon Brinkmann$^{11}$,   Benjamin A. Weaver$^{9}$ \\
\\
\\
$^1$Institute for Cosmology and Gravitation, University of Portsmouth, Portsmouth, PO1 3FX, UK \\
$^2$SEPnet, South East Physics Network, ({\tt www.sepnet.ac.uk})\\
$^3$Department of Astronomy, University of California, Berkeley, CA, USA \\
$^4$Lawrence Berkeley National Lab, 1 Cyclotron Road, Berkeley, CA 94720, USA\\
$^5$Berkeley Center for Cosmological Physics, University of California, Berkeley, CA 94720, USA\\
$^6$Department of Astrophysical Sciences, Princeton University, Peyton Hall, Princeton, NJ 08544, USA \\
$^7$Department of Astronomy and Astrophysics, The Pennsylvania State University, 525 Davey 
Lab, University Park, PA 16802, USA \\
$^8$Steward Observatory, University of Arizona, 933 N. Cherry Ave.
Tucson, AZ 85721, USA \\
$^9$Center for Cosmology and Particle Physics, New York University, New York, NY 10003, USA\\
$^{10}$Department of Astronomy, Yale University, New Haven, CT 06520, USA\\
$^{11}$Apache Point Observatory, Apache Point Road, P.O. Box 59, Sunspot, NM 88349, USA 
\\ 
\\
{\tt E-mail: karen.masters@port.ac.uk}
 }

\maketitle

\begin{abstract}
We study the morphology and size of the luminous and massive galaxies at
$0.3< z < 0.7$ targeted in the Baryon Oscillation Spectroscopic Survey
(BOSS) using publicly available Hubble Space Telescope (HST) imaging,
and catalogues, from the COSMic Origins Survey (COSMOS). Our sample (240
objects) provides a unique opportunity to check the visual morphology of
these galaxies which were targeted based solely on stellar population
modeling. We find that the majority of BOSS galaxies (74$\pm6$\%)
possess an early-type morphology (elliptical or lenticular), while the
remainder have a late-type (spiral disc) morphology. This is as expected from the goals
of the BOSS target selection which aimed to predominantly select
slowly evolving galaxies, for use as cosmological probes, while still
obtaining a fair fraction of actively star forming galaxies for galaxy evolution
studies. We show that a colour cut of $(g-i)>2.35$ is able to select a
sub-sample of BOSS galaxies with $\ge 90$\% early-type morphology and
thus more comparable to the earlier Luminous Red Galaxy (LRG) samples of
SDSS-I/II. The remaining $\simeq10$\% of galaxies above this $(g-i)$ cut
have a late-type morphology and may be analogous to the ``passive
spirals" found at lower redshift. We find that 23$\pm4$\% of the
early-type BOSS galaxies are unresolved multiple systems in the SDSS
imaging. We estimate that at least 50\% of these multiples are likely real associations and not projection effects and may represent a significant ``dry merger" fraction.   We study the SDSS pipeline sizes of BOSS galaxies which we find to be systematically larger (by 40\%) than those measured from HST images, and provide a statistical correction for the difference. These details
of the BOSS galaxies will help users of the BOSS data fine-tune their
selection criteria, dependent on their science applications. For
example, the main goal of BOSS is to measure the cosmic distance scale
and expansion rate of the Universe to percent-level precision -- a point
where systematic effects due to the details of target selection may
become important.

\end{abstract}


\section{Introduction}

The Baryon Oscillation Spectroscopic Survey (BOSS) is one of four surveys being carried out as part of the Sloan Digital Sky Survey III (SDSS-III; Eisenstein et al. 2011). The main goal of BOSS is to measure the cosmic distance scale and expansion rate of the Universe with percent-level precision at $z < 0.7$ and $z\sim2.5$ using the Baryon Acoustic Oscillation (BAO) scale. To achieve this goal, BOSS is performing a redshift survey of 1.5 million massive luminous galaxies (between $0.3<z<0.7$) and 150,000 quasars at $z>2.5$ selected from the SDSS imaging data published as part of Data Release Eight (DR8; Aihara et al. 2011). 
 
Previous cosmological surveys of distant luminous galaxies have focused on the reddest galaxies because they are efficient tracers of the underlying dark matter distribution, e.g. Luminous Red Galaxies (LRGs; Eisenstein et al. 2001, Cannon et al. 2006). However, BOSS is now including bluer galaxies than these previous samples partly to increase the sky density of sources observed (to improve the BAO measurements), as well as to sample intrinsically bluer galaxies to provide a more representative census of the galaxy population for comparison with models of galaxy evolution (e.g. \citealt{K93}).  
 
With the goal of percent-level precision on the cosmological parameters, BOSS is entering the regime where the details of galaxy evolution, and sample selection, play an important role in understanding the systematic errors associated with accurate measurements of the clustering of galaxies on large scales. Different types of galaxies populate their dark matter halos differently (e.g. Hogg et al 2003, Zehavi et al. 2005, Ross \& Brunner 2009), and especially evolve on different timescales so a detailed understanding of the galaxy types in the BOSS sample is important both for cosmology and studies of galaxy evolution.  

Galaxies are selected for inclusion in BOSS (and many other modern surveys) using stellar population models which predict the expected colours as a function of the galaxy star formation history, dust content and redshift. This method is effective as the stellar evolutionary status of a galaxy imprints onto its spectral energy distribution, hence colours, with the well-known effect that galaxies dominated by old stars, or more generally having a small scatter in age - are redder than galaxies hosting star formation. To a large extent this distinction is mirrored in the galaxy morphology, with older and redder galaxies typically having early-type morphologies, while star-forming galaxies usually possess discs. However, since a galaxy's morphology is also affected by gas accretion and merging histories which effect the dynamics of the stars, the colour and morphology of a galaxy can become decoupled. 

In recent years, there has been a resurgence of interest in the morphology of galaxies, especially at low redshift. For example, Schawinski et al (2007) and Thomas et al (2010) constructed a sample of 50,000 visually--inspected early-type galaxies from SDSS, and showed that morphology was an important factor in the interpretation of the transition region between the ``blue cloud" (late--type galaxies) and the ``red sequence" (early--type galaxies). These studies discovered and discussed so-called ``blue ellipticals", i.e. galaxies with an elliptical morphology and blue colours \citep[also confirmed by][]{S09, kap2009}. The Galaxy Zoo project (Lintott et al. 2008), with the help of citizen scientists have now provided visual classifications for all galaxies in the original ($z<0.3$) SDSS-I/II data (Lintott et al. 2011). This work has confirmed the strong correlation between colour and morphology of galaxies, but in addition hi-lighted objects in which the colour and morphology have become decoupled, such as the blue ellipticals described above, and also a population of ``red spirals" \citep{B09,M10}, which have also been found elsewhere \citep[e.g.][]{W09, B10} and represent a significant fraction of the massive spiral population, as well as possibly an important evolutionary stepping stone for all massive late-types moving to the red sequence. Moreover, such work has re-defined the classic morphology-density relation (Dressler et al. 1980) at low redshift as predominantly a colour-density relation, i.e., at a fixed galaxy colour, there is a weak morphology--density relationship, while at a fixed morphology, there remains a significant range of colours with density (see Blanton et al. 2005, Ball, Loveday \& Brunner 2008; Bamford et al. 2009; Skibba et al. 2009). 
      
In this paper we shall study the visual morphology, and catalogued sizes, of galaxies in the BOSS survey which have been targeted using colours predicted by population models. We shall accomplish the double aim of checking the effectiveness of a purely colour-based selection in targeting mostly passive early-type galaxies as well as determining the relation between colour and morphology at these redshifts. With a median seeing of 1.1\arcsec ~for the SDSS DR8 imaging data, the BOSS target galaxies are barely resolved (see Figures 2, 3 \& 5) thus making it difficult to obtain reliable visual morphologies or sizes from these data. Therefore, higher angular resolution (and deeper) imaging is needed to determine the detailed shapes of BOSS galaxies, and the best facility currently available for studying such high-redshift galaxy populations is the Hubble Space Telescope (HST).  

 We construct a sample of BOSS targets detected in the publicly available HST imaging of the COSMic Origins Survey (COSMOS; Scoville et al. 2007). Although small (240 objects), such an unbiased match between COSMOS and BOSS provides an important sample of objects for statistical studies and is unlikely to grow in size or depth over the next few years. 

Where appropriate, we assume a cosmological model of a flat, $\Lambda$CDM Universe with $H_0=70$ \kmsMpc~ and $\Omega_M = 0.3$ consistent with recent WMAP observations \citep{komatsu11}.
    
\section{Sample and Data}
The SDSS-III DR8 imaging data covers 15,000 deg$^2$ of high galactic latitude sky comprising of the SDSS-I/II Legacy survey and 3000 deg$^2$ of new imaging concentrated in the Southern Hemisphere \citep{Y00,DR8paper}. Images in five filters ($ugriz$, \citealt{F96}) are obtained with a CCD camera \citep{G98} mounted on the SDSS Telescope \citep{G06}. 

Spectroscopic targets for BOSS are selected from the high Galactic latitude regions of this imaging (around 10,000 deg$^2$ in total) using colour cuts designed to select luminous, massive galaxies with an approximately uniform distribution in stellar mass over the redshift range $0.3<z<0.7$. The details of target selection are given in Eisenstein et al. (2011) and summarised in Section \ref{target} below. The BOSS galaxy sample is designed to have a total angular source density of approximately 150 galaxies deg$^{-2}$  which is an order of magnitude greater than the SDSS-I/II LRG sample.

\subsection{COSMOS HST Imaging}

COSMOS is the largest HST survey yet undertaken and was specifically designed to survey all types of galaxy environments at $z>0.5$ (Scoville et al. 2007). The HST imaging for COSMOS \citep{K07,L07} is publicly available\footnote{See {\tt http://irsa.ipac.caltech.edu}} over a 1.64 deg$^2$ field centred at $\alpha=10^h$, $\delta = 2.2^\circ$ (J2000). The COSMOS field is covered by the SDSS-I/II Legacy imaging data and in total, 240 BOSS targets are found in common between the two sets of images, consistent with the expected source density of BOSS targets (\ie. 240 targets over $1.68$ deg$^2$ = 143 galaxies deg$^{-2}$). This COSMOS-BOSS sample can be considered largely representative of the entire BOSS galaxy sample, particularly in the higher redshift CMASS sub-sample (see Section 2.3), but we note that in the lower redshift LOZ subset ($0.2<z<0.4$), the presence of a significant over-density in the COSMOS field at $z=0.35$ \citep[e.g.][]{cosmosdensity} may bias the morphological mix towards red, early types through the morphology-density relation. 

\begin{figure}
\includegraphics[width=8.4cm]{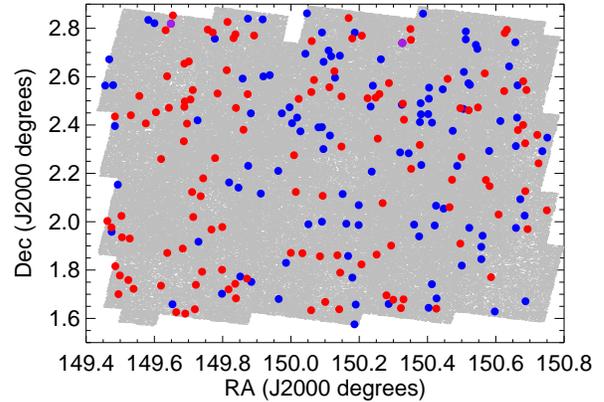}
\caption{The sky distribution of BOSS target galaxies in the COSMOS field is shown as coloured points over plotted on the positions of all galaxies detected by COSMOS (small grey points). We identify BOSS targets by their sub-sample (see Section 2.3); CMASS objects are shown in red, while LOZ objects are shown in blue. Two of the targets fit both target criteria and are shown as purple.  
\label{sky}}
\end{figure}

We show the sky distribution of these 240 BOSS targets compared to all COSMOS identified galaxies in Figure \ref{sky}. For these 240 BOSS targets, we have visually inspected images taken by the Advanced Camera for Surveys (ACS) instrument on HST in the I-band (F814W) filter, which at the typical redshift for BOSS, corresponds to a rest--frame V-band filter. We also inspect
I- and V-band colour composite images provided by Griffith et al. (in prep.)\footnote{{\tt http://www.ugastro.berkeley.edu/$\sim$rgriffit/Morphologies/}} and a different set of colour composite images made for the Galaxy Zoo: Hubble project (Lintott et al. in prep).

We note that ACS images of other BOSS galaxies will be available in the HST public archive in images taken for other observations. 
A preliminary search of the HST ACS public archive finds 795 images to the same depth as the COSMOS data in the F814W filter. Of these, only 563 were unique (i.e., the centres of the pointings are separated by more than 10 arcseconds), and 398 of these 563 images are within the approximate SDSS-III BOSS imaging area. This is less than the 567 ACS images which make up COSMOS, and is roughly 70\% of the area of COSMOS so would add a likely $\simeq 170$ more galaxies to our comparison sample (assuming the same surface density of BOSS galaxies). While this figure could increase the sample size of our study, we note that these additional ACS fields, and associated galaxies, could be potentially biased in their morphological mix, e.g., if found in HST imaging of the cores of groups and clusters of galaxies, whereas the COSMOS field would specifically avoid such bias. We also make use of the publicly available Zurich Estimator of Structural Types (ZEST) measurements \citep{ZEST} for our BOSS galaxies in COSMOS and such measurements have not been made for the entire HST archive.
Hence, we decide to rely only on BOSS galaxies imaged by HST in the COSMOS area. 

\subsection{BOSS Imaging and Spectra}

Throughout this paper, we use the SDSS-III DR8 pipeline photometric measurements as described in \cite{DR8paper}. Information on the SDSS pipeline and algorithms is also given in \cite{S02}. 

For the colours of BOSS galaxies, we use the model magnitudes. In the SDSS pipeline \citep{S02}, both a de Vaucouleurs and an exponential profile fit is performed on each galaxy, and the best fit model is a linear combination of the two profiles. Model magnitudes are calculated using the profile shape fit in $r$-band and applied to each of the five SDSS bands (allowing for a variable amplitude). For apparent luminosities, we use cmodel magnitudes where the model is fit individually in each SDSS band. All magnitudes are corrected for Galactic extinction using reddening estimates from the DIRBE maps by Schlegel, Finkbeiner \& Davis (1998). 

For SDSS measured galaxy sizes, we use the effective radius from the de Vaucouleur fit. This is a seeing--corrected quantity, albeit with some approximations made, so seeing may have a residual effect as we will discuss further in Section 4, and especially because the typical sizes of BOSS galaxies are comparable to the size of the seeing disk. 

 The SDSS seeing in the COSMOS field is typical for the whole DR8 imaging, e.g., the median seeing for the whole SDSS is 1.08\arcsec ~with an interquartile range of 0.95-1.22\arcsec ~(Ross et al. 2011)\footnote{The improvement in seeing over previous SDSS imaging releases is due to the large fraction of repeat images that have been taken. In such cases, the best seeing is selected for release, thus reducing the median seeing level significantly over earlier data releases.}. The median seeing for our subset of BOSS targets in COSMOS is 1.10\arcsec, with an interquartile range of 1.02-1.16\arcsec ~(we are quoting the parameter {\tt PSF\_FWHM} in $i$-band from DR8). 

An important caveat to this work is that the SDSS imaging in the COSMOS field has a higher sky noise level than most of the other DR8 imaging.
The median sky noise level in the whole DR8 is 20.27 mags (Ross et al. 2011)\footnote{Ross et al. (2011) report this value expressed in ``nanomaggies", which are a linear unit of flux used in the SDSS collaboration. We use the conversion from a flux, $f$, in nanomaggies to, $m$ in magnitudes of $m= 22.5 ~{\rm mag} - 2.5 \log f$}, with an upper quartile of 20.11 mags. Almost all (97\%) of the BOSS targets in the COSMOS imaging have SDSS sky noise values in this upper quartile, which of course translates to noisier photometry, and shallower images than are typical for DR8. This may have an effect, in particular, on the SDSS measured sizes. 

BOSS spectra were taken for 171 of the BOSS galaxy targets on 10th March 2011 (note that the entire COSMOS area fits within a single SDSS plate). These observations resulted in 166 BOSS galaxy redshifts. The observations which did not result in galaxy redshifts are made up of 2 spectra which had unreliable fits, or {\tt zwarning = 4}; and 3 spectra of stars in our Galaxy. An additional 58 of the targets in our sample had redshift measurements which were taken as part of the SDSSI/II program. The remaining 11 targets could not be observed due to problems with fiber collisions.  In total this adds up to 224 of the BOSS targets (93\%) having reliable galaxy redshifts. Where appropriate, we use the redshift measurement for these galaxies, but do not use any additional information from the BOSS (or SDSS) spectra. We defer such an analysis for future work. 

\subsection{BOSS Target Selection \label{target}}

The BOSS galaxy target selection is based on colour and apparent magnitude cuts defined in the observed-frame $g,r,i$ photometry which have been derived via population synthesis models. It also includes cuts based on PSF magnitudes which are used to reduce stellar contamination. The exact cuts are given in Eisenstein et al. (2011). For the scope of this paper, it suffices to recall that the BOSS target selections are split into two sub-samples which are aimed at selecting galaxies in the different redshift ranges of BOSS, namely:  (1) a Low Redshift Sample (LOZ), aimed at selecting luminous massive galaxies at $0.2<z<0.4$ (see the area below the dashed line in Figure 4); and (2) a higher redshift Constant Mass Sample (CMASS), selecting luminous and massive galaxies at $z>0.4$ (area above the dashed line in Figure 4). Note that the CMASS, high-$z$ cut is a new BOSS cut, while the LOZ cut is similar to the original Luminous Red Galaxy (LRG) Cut I selection from SDSS-I/II (Eisenstein et al. 2001).

The BOSS target selection cuts are based on the expected track of a passively--evolving, constant stellar mass galaxy based on a population model for luminous red galaxies by \citet{M09}. As has been discussed previously in both Eisenstein et al. (2011) and White et al. (2011), the CMASS colour selection closely tracks the redshifted colours of luminous and massive galaxies at $z\simeq0.5$, while the (colour dependent) $i$-band magnitude constraint in CMASS is aimed at selecting objects with $M_\star \simeq 10^{11} M_\odot$, independent of their intrinsic rest-frame colours.  Figure 1 of \citet{White11} demonstrates that this set of cuts successfully isolates galaxies at $z>0.4$.

What is most relevant to our study is that: i) the main colour cut for CMASS (the selection of $d_\perp = (r-i) - (g-r)/8.0 > 0.55$) is basically a selection on the observed-frame colour $(r-i)$ (\ie~ the line is almost horizontal in Figure 4). All galaxies in the redshift range of CMASS ($0.4<z<0.7$) have their $4000\AA$ break feature somewhere in the $r$-band since at $z=0.4$ the break is exactly between $g$ and $r$ at $5600\AA$, and at $z=0.75$ it moves into $i$-band (at $7000\AA$). This means that all galaxies at these redshifts are red in observed frame $(r-i)$, regardless of stellar content; ii) no strong cut was made in the observed-frame $(g-r)$, which would select approximately in intrinsic $(u-g)$ colour at these redshifts -- a colour which is very sensitive to stellar content; iii) a (slightly colour dependent) $i$-band magnitude based cuts is applied to require that (only) massive galaxies are selected. 

The stellar population model based target selection for BOSS described above, implies that we should find within the sample mostly early-type galaxies -- since this type of galaxy dominates the massive galaxy population.  But, given the absence of a red colour cut in $(g-r)$, massive blue galaxies could also be found. We shall see in the next Section (Figure 4) that the real morphologies of the BOSS galaxies in our sample reflect well this expectation.

\section{Visual Morphology}

We visually inspected both the COSMOS and SDSS images of all \nmatch BOSS target galaxies in the COSMOS area. The COSMOS images were used to first classify galaxies as either ``early" or ``late-type" morphology. We then further separated these broad classes into possibly lenticular (S0) early-type galaxies, disturbed galaxies, and/or multiple systems (including multiple cores). For the late-type galaxies, we also recorded if the galaxy was viewed as ``edge-on" disc or possessed a bar through the middle of the galaxy.

\begin{figure*}
\includegraphics[width=15cm]{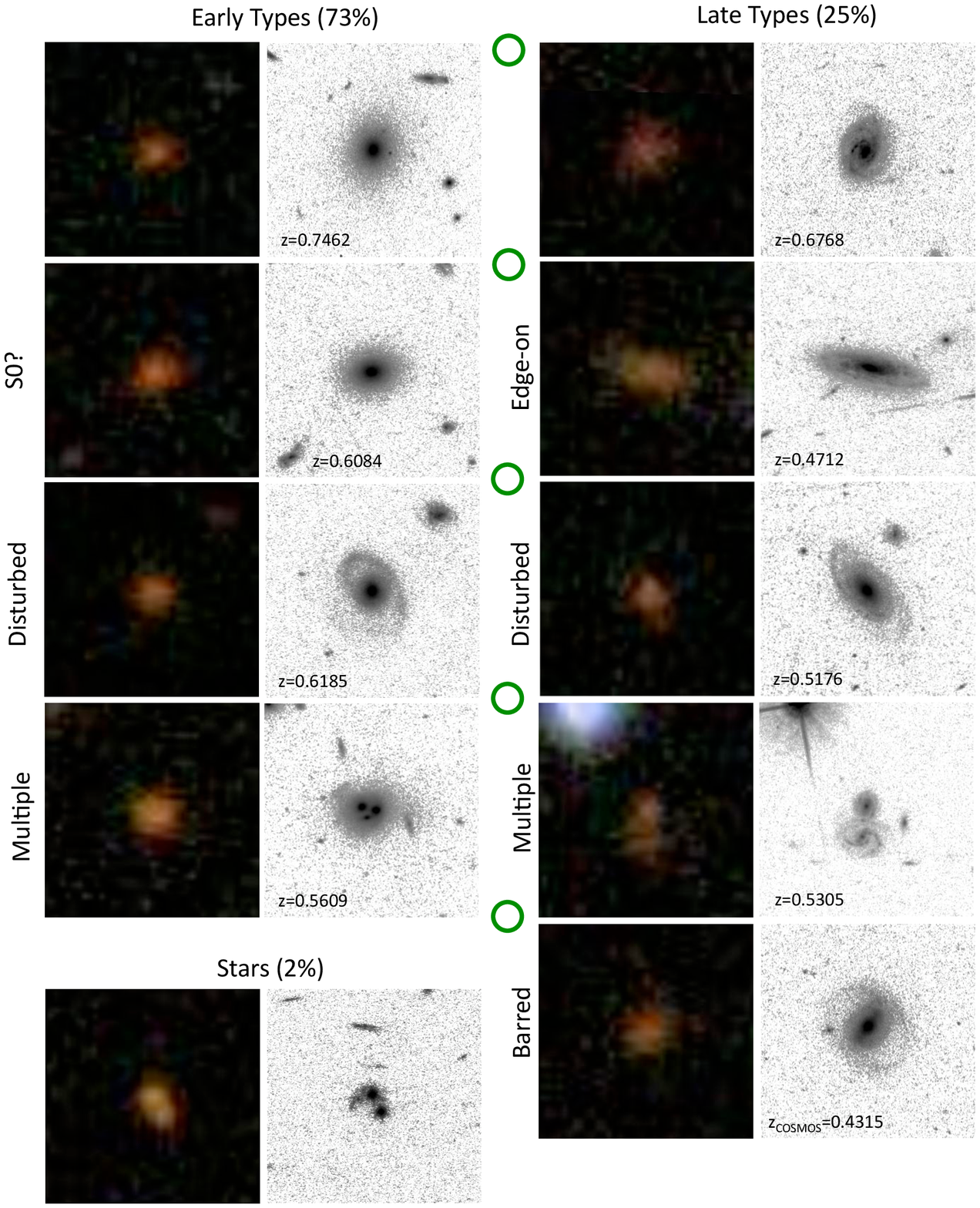}
\caption{Examples of SDSS composite colour $gri$ images and ACS I-band images (shown as black on white) for BOSS CMASS galaxies with different morphological types. Shown are 10 examples out of a total sample of 129 objects. All images are 15\arcsec ~square, and the BOSS 2\arcsec ~diameter fibre is illustrated by a green circle. The redshifts of these objects are shown in the figure. The barred late--type was not observed by BOSS due to fibre collisions, but has a redshift from zCOSMOS (Lilly et al. 2007). The double point source system has also been observed spectroscopically by BOSS and is confirmed to be made up of stars in our Galaxy (\ie ~$z=0$). 
\label{morphfig}}
\end{figure*}

\begin{figure*}
\includegraphics[width=15cm]{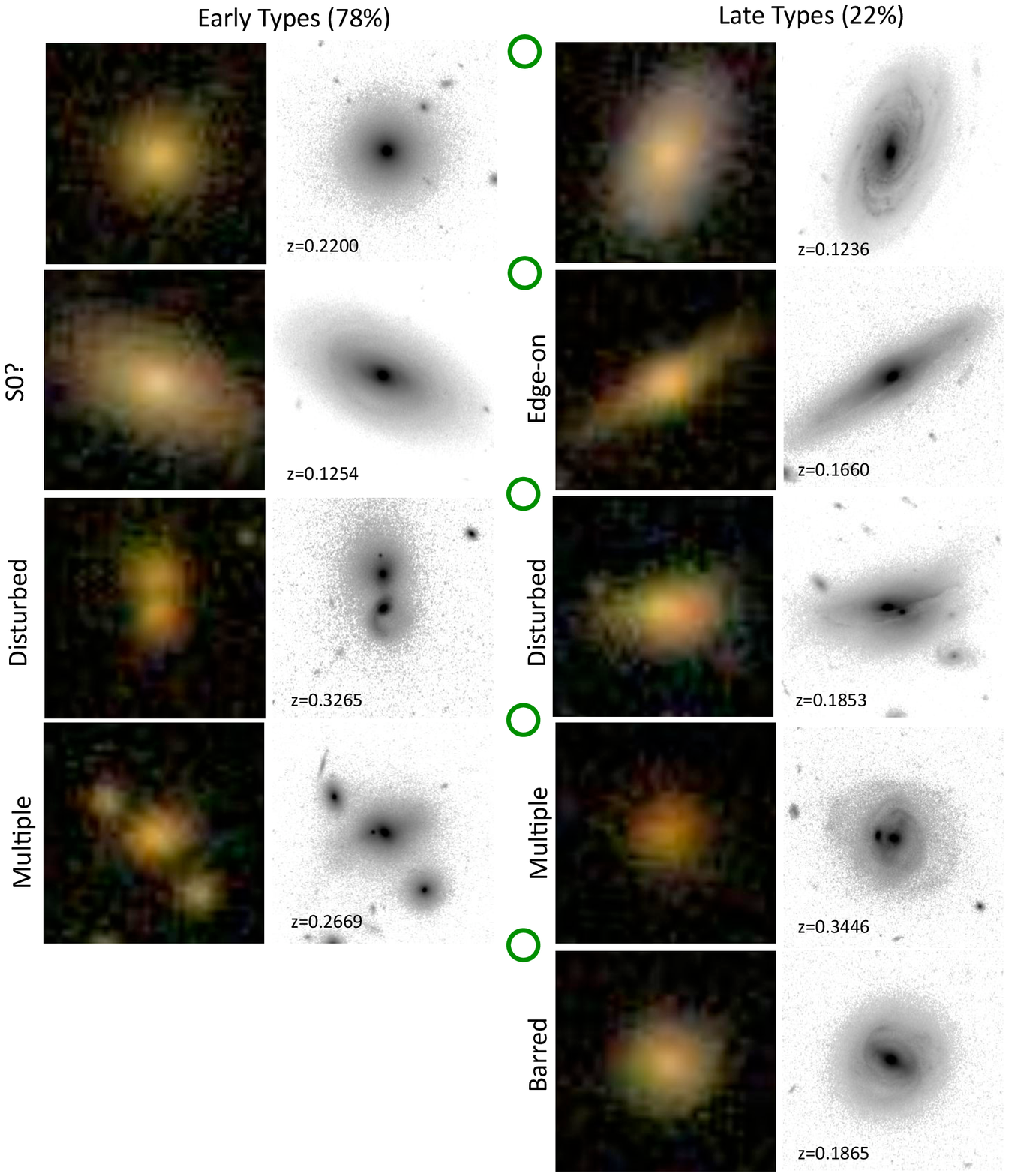}
\caption{As with Figure \ref{morphfig} but for the BOSS LOZ sample which has a total of 113 objects (and no point sources). The redshifts of these objects mostly come from SDSS-I/II observations). We note that the multiple late-type galaxy also makes the colour cuts of the BOSS CMASS sample.  
\label{LOZmorphfig}}
\end{figure*}

Table \ref{morph} shows the morphological mix of our sample, and we present all of our visual classifications in Appendix A. We find that the majority of BOSS galaxies in our COSMOS sample are early-type galaxies ($74\pm6\%$; we quote the Poisson error on fractions throughout, but note that this is strictly only valid for large sample sizes and fractions neither close to 0\% or 100\%. The error we quote will therefore be an underestimate of the true confidence region particularly for the small fractions and subsets; see \citep{C11} for more details), but there is a substantial fraction of late-type galaxies present (24$\pm3\%$). Representative images from each morphological class of object are provided in Figure \ref{morphfig} for the CMASS sample and in Figure \ref{LOZmorphfig} for the LOZ sample, and all 240 galaxies are available for inspection at {\tt http://icg.port.ac.uk/$\sim$mastersk/BOSSmorphologies/} which shows both the SDSS $gri$ colour composites alongside the COSMOS HST I-band and colour composite images.

\begin{table*}
\caption{Visual morphology of BOSS galaxies in the COSMOS imaging}
\label{morph}
\begin{tabular}{lccccc}
\hline
Type  & All BOSS & CMASS  & LOZ  & $(g-i)>2.35$ & CMASS  $(g-i)>2.35$ \\
\hline
All objects            & 240               & 129           & 113           & 135               & 91\\
\hline
Early-types &     178 (74\%) &      91 (71\%) &      88 (78\%) &     122 (90\%) &      83 (91\%) \\
~~~~Lenticular  &      ~34 (19\%) &      15 (16\%) &      19 (22\%) &      ~20 (16\%) &      12 (14\%) \\
~~~~Multiple    &      ~40 (23\%) &      21 (23\%) &      20 (23\%) &      ~27 (22\%) &      19 (23\%) \\
~~~~Disturbed   &      ~26 (15\%) &      14 (15\%) &      12 (14\%) &      ~17 (14\%) &      12 (14\%) \\
\hline
Late-types  &      58 (24\%) &      34 (26\%) &      25 (22\%) &      13 (10\%) &       8 ( 9\%) \\
~~~~Barred      &      11 (28\%) &       ~8 (29\%) &       ~4 (33\%) &       4 (57\%) &       4 (57\%) \\
~~~~Edge        &      18 (31\%) &       ~6 (18\%) &      12 (48\%) &       6 (46\%) &       1 (13\%) \\
~~~~Multiple    &       ~6 (10\%) &       ~5 (15\%) &       ~2 ( 8\%) &       2 (15\%) &       2 (25\%) \\
~~~~Disturbed   &      13 (22\%) &      11 (32\%) &       ~3 (12\%) &       4 (31\%) &       4 (50\%) \\
\hline
Stars & 4 (2\%) & 4 (2\%) & - & - & -\\
~~~~Multiple & 2 & 2 & - & -& -\\
\hline
\end{tabular}
\end{table*}

\subsection{Morphology and colour}

 \begin{figure*}
\includegraphics{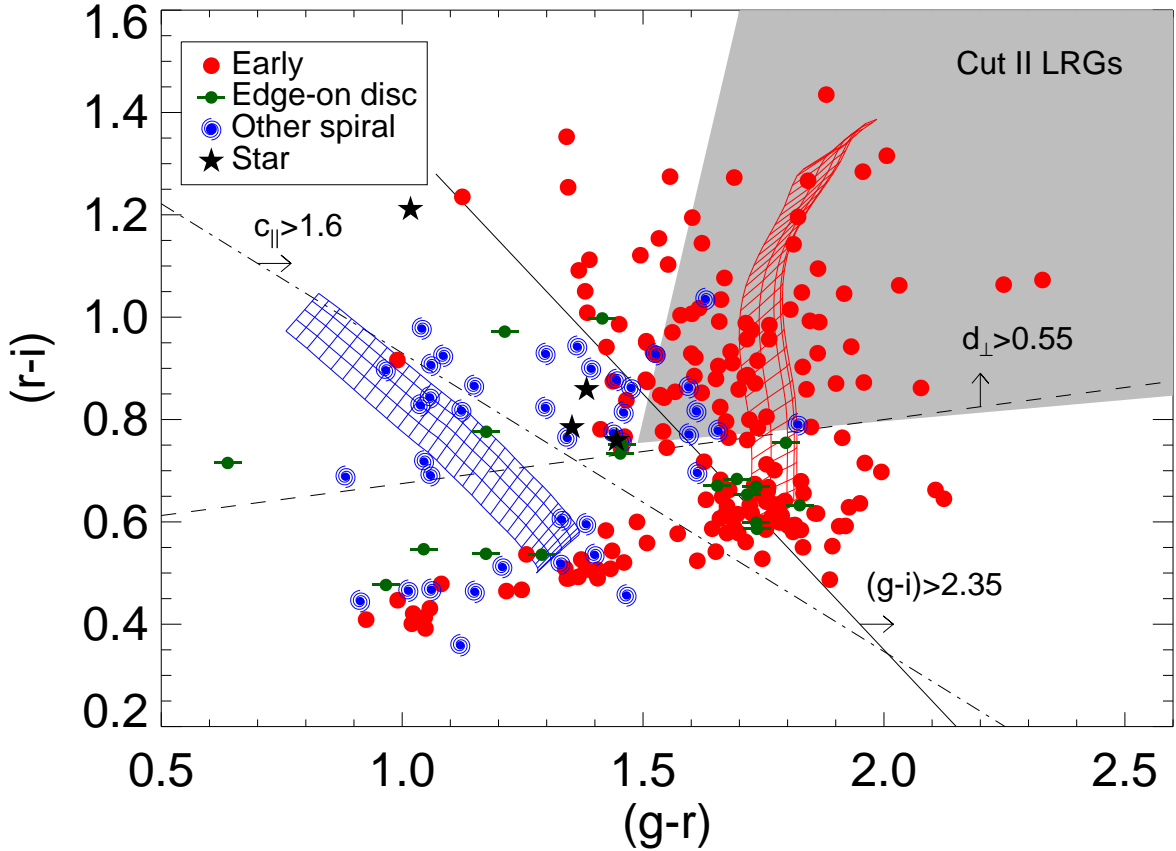}
\caption{The $(g-r)$ versus $(r-i)$ colours for our sample of BOSS galaxies found in COSMOS HST imaging. The median photometric errors on these colours are $\sigma_{(r-i)}=0.05$ and $\sigma_{(g-r)}=0.11$). The dashed line shows the $gri$ colour combination, $d_\perp>0.55$ limit for the CMASS sample (CMASS objects lie above this line). 
The solid black line illustrates a simple colour cut at $(g-i)=2.35$ which can be used to remove most of the late-types in the BOSS sample. The dot-dashed line shows the $gri$ colour combination, $c_{||}>1.6$ cut used by the 2SLAQ survey (Cannon et al. 2006) to define LRGs. The shaded region in the top right--hand corner of the plot shows the Cut II LRG sample from the original SDSS-I/II (Eisenstein et al. 2001). The red and blue grids show theoretical model tracks for passive LRGs (red) and constantly star-forming galaxies (blue) with $0.4<z<0.7$ (grid shows $\Delta z=0.01$) and ages of 4, 5, 6, 7  and 8 Gyrs. 
\label{colour}}
\end{figure*}

BOSS target galaxies in the COSMOS field are shown in Figure \ref{colour} as a function of their observed frame $(g-r)$ and $(r-i)$ SDSS colours\footnote{Colours include Galactic extinction corrections}. The symbols indicate the basic visual morphology of the galaxies, while the dashed diagonal line is the $d_\perp$ limit of the CMASS sample, i.e. galaxies above this line are in the CMASS subsample and should have redshifts larger than $z=0.4$. 

We see that the majority of objects both above and below the CMASS cut, have an early-type morphology (shown as red circles) and that late-type galaxies (blue spiral symbols) in BOSS lie mostly at the bluest $(g-r)$ colours.

We suggest a simple colour cut of $(g-i)=2.35$ above which almost all (90\%) of our BOSS galaxies have an early-type morphology in both the LOZ and CMASS samples (numbers are given in Table 1). Even possibly dust--reddened inclined spirals are rarely redder (in the optical) than this colour cut. For example, for the CMASS sample, 91\% of the early-type galaxies are redder than $(g-i)=2.35$ while 76\% of the late-type galaxies are bluer than this cut. 
 
The power of such a simple $(g-i)$ colour cut can be easily understood by looking at Figure 4, where theoretical model tracks (from Maraston 2005 and Maraston et al. 2009) for different star formation histories, dust content and age are plotted in the observed-frame for redshifts $0.4<z<0.7$. The age of the models, intended as the time after the start of star formation, is between 3 and 8 Gyr, which covers the age of the Universe adopting a galaxy formation redshift of $z=5$. The red grid shows the LRG model of Maraston et al. (2009), which nicely covers the area where the CMASS early-types are found. The bluest late-types within CMASS (at $(g-r)\sim1$~and $(r-i)\sim0.8$) are well represented by a model with constant star formation ($\tau\sim 20$ Gyr)~and a E(B-V)=0.3 using a Calzetti et al. (2000) reddening law (blue grid). 

 We point out that at the CMASS redshifts, a $(g-i)$ observed frame colour is almost equivalent to a $(u-r)$ rest frame colour (the centre of $u$ shifts to $g$ at $z=0.4$; while $r$ shifts to $i$ at $z>0.2$). Our $(g-i)>2.35$ colour separation can therefore be compared to $(u-r)$ colour selections in the local galaxy population, so we note that this cut is very similar to the optimal separator of $(u-r)=2.22$ derived by Strateva et al. (2001) for the SDSS main galaxy sample (MGS), and to the similar $(u-r)$ cut derived as a function of $M_r$ by Baldry et al. (2004) which asymptotes to roughly $(u-r)=2.3$ for the most luminous galaxies.  

Cannon et al. (2006) used a similar colour cut ($c_{||}>1.6$; which we show in Figure \ref{colour} by a dot-dashed line) which was introduced to remove late--type galaxies in the 2SLAQ LRG sample. This colour cut is often used in photometric redshift studies (e.g. Ross et al. 2011) as it is has been empirically observed to help remove galaxies with unreliable photometric redshifts. As is clear in Figure 4, our proposed colour cut is more efficient in removing late-type systems particularly in the CMASS galaxy sample. 

We also show in Figure \ref{colour} the Cut II LRG selection from SDSS-I/II \citep{E01} (shaded region). In the COSMOS area, 70 BOSS targets make this colour selection, with 91\% ($64$) of them being early-type galaxies. Note that the BOSS sample is deeper than the original Cut II LRG selection which had an average source density of 12 LRGs deg$^{-2}$. A total of only four original SDSS-I/II LRGs were observed in the COSMOS field. Finally, we remind the reader that the LOZ selection of BOSS (points with $d_\perp<0.55$ in Figure \ref{colour}) is similar to (but deeper than) the SDSSI/II Cut I LRG sample. A total of 51 Cut I LRGs were observed in the COSMOS area. 

In the following sections we give more details on the various morphological sub-types.
  
\subsection{Unresolved Objects}

We find four of the 240 BOSS targets which in the HST imaging are made up of unresolved point sources (stars or AGN). These objects passed the BOSS star--galaxy separation, but clearly have diffraction spikes in the COSMOS imaging. Two of the four point source systems are actually composed of two point sources each which are merged into single objects in BOSS imaging. We expect that the COSMOS field might have a lower stellar contamination rate than the rest of the survey because it was selected as a deep extragalactic field and is located away from the plane of our Galaxy at a Galactic latitude of $b=42^\circ$. However, our visual inspection gives a point source contamination rate of $2\pm1$\% for BOSS targets which agrees, within the error, with the stellar contamination rate from the BOSS spectra taken to date, i.e., 3.38\% of BOSS spectra taken of targeted galaxies are stars (David Wake, private comm.). 
 
 In addition we look at objects in the COSMOS field which pass all BOSS target selection except the star--galaxy separation (which is based on the difference between PSF and model magnitudes in the SDSS pipeline). For the LOZ subset there are three such objects, all of which are point sources in the HST imaging. In the CMASS subset there are 18 such objects, four of which turn out to be galaxies (extended objects) in the HST imaging. This indicates that 3$\pm2$\% (4/129) of CMASS galaxies or 2$\pm1$\% (4/240) of all BOSS target galaxies in COSMOS are missed due to the star-galaxy separation, consistent with our expectation of $<2$\% for the whole survey.
  
\subsection{Late-type Galaxies}
     
Of the late-type galaxies in the BOSS COSMOS sample  (which make up $24\pm3$\% of all BOSS target galaxies), approximately a third are viewed edge-on (31$\pm8$\%), by which we mean they show the classic Sp, or ``spindle" spiral type where no part of the face-on disc can be seen -- we estimate this includes objects within about 5--10$^\circ$ of completely edge-on. This value is larger than the fraction expected from random inclinations on the sky, where we would expect for example, only 10\% of the systems should be within 5$^\circ$ of edge-on, or a quarter within $15^\circ$ of edge-on.

 Breaking the late-types into CMASS and LOZ subsamples, we find that the bulk of this excess of edge-on disks is caused by the LOZ subsample, with the CMASS sample having an edge-on fraction consistent with random (6/34 objects, or 18$\pm$7\%), while edge-on disks in the LOZ sample make up almost half of the late-types (12/25 objects, or 48$\pm$14\%). 

 The high fraction of edge-on disc galaxies in the LOZ subsample suggests that something in the LOZ target selection is preferentially allowing in disc galaxies if they are seen in the edge-on orientation. Edge-on disks will be observed to be both redder (due to dust reddening) and have higher surface brightness and concentration than if seen face-on  (e.g. Masters et al. 2010a). As we discuss above, the LOZ subsample is a fainter version of the SDSS I/II Cut I LRG selection (Eisenstein et al. 2011). Unlike the CMASS selection (which has a colour cut designed to select only on redshift) this LRG selection for LOZ has a cut on intrinsic colours which risks preferentially allowing in edge-on dust reddened discs. For example there is a cluster of edge-on spirals at colours of $(g-r)>1.7$ and $(r-i)\simeq0.6$ in Figure \ref{colour}). We expect that surface brightness selection is not likely to be a big issue for BOSS late-type galaxies which are all massive spirals and as such should not even when seen face-on have low enough surface brightness to miss selection. 
  

Overall, 8$\pm2$\% of all BOSS targets studied here ($5\pm2$\% of CMASS and 12$\pm3$\% of LOZ) are highly inclined disc galaxies, and could represent a worrying contamination to the BOSS samples. For example, Yip et al. (2011) and Ross et al. (2011) have investigated inclination--dependent reddening as a source of systematic error for photometric redshift estimates. Also, since the intrinsic rest-frame colours of these edge-on disc galaxies are almost certainly bluer than their ``observed" rest-frame colours, and their luminosities are dimmed, their true stellar masses may be biased from estimated values. The amount, and sense of this stellar mass bias depends on the details of the colours and luminosity used in a stellar mass estimate \citep{Ma09}. Usually, the stellar masses derived from full SED fitting of highly-reddened galaxies with on-going star formation are underestimated because of the age-dust degeneracy \citep{CM10}. In some stellar mass estimates, the reddening and dimming can cancel out the effect on the stellar mass (e.g. in the case of the the colour-dependent $M/L$ relation provided by \citealt{B01} dust attenuation moves galaxies parallel to the relation), but the cancellation is often only partial \citep[e.g.][]{R11}, so care still needs to be taken. 

 Reassuringly, for many applications which will use only the CMASS subset, e.g. Ross et al. (2011), most of the edge-on discs are not found in the CMASS cut. Moreover, note that no inclined disc galaxy is as red as the reddest early--type galaxies in our sample of BOSS galaxies (e.g., towards the top right--hand corner of Figure \ref{colour}). This result suggests that dust reddening, even if significant, is a smaller effect than the natural spread in optical colours of massive galaxies due to their stellar populations combined with redshift.  A similar result was found at lower redshift for the SDSS-I/II Galaxy Zoo sample (see Masters et al. 2010a).
  
We find that some of the non-inclined late-type galaxies in BOSS are still as red as the early-type galaxies, i.e., 10 $\pm 4$\% of the galaxies redder than our proposed $(g-i)=2.35$ colour cut in the CMASS sample (see Section 3.1 for details) are non-inclined late-type galaxies. These may be the high-redshift counterparts of the \klm{passive, red} spirals seen at lower redshifts in the SDSS Main Galaxy Sample \citep{M10}. Red disc galaxies have already been identified in the COSMOS imaging, at approximately the same fraction; e.g., \citet{B10} found $15\%$ of red sequence objects had discs at $z\simeq0.5$. 

\subsection{Pairs, Multiples and Disturbed Systems}
\begin{figure*}
\includegraphics[width=15cm]{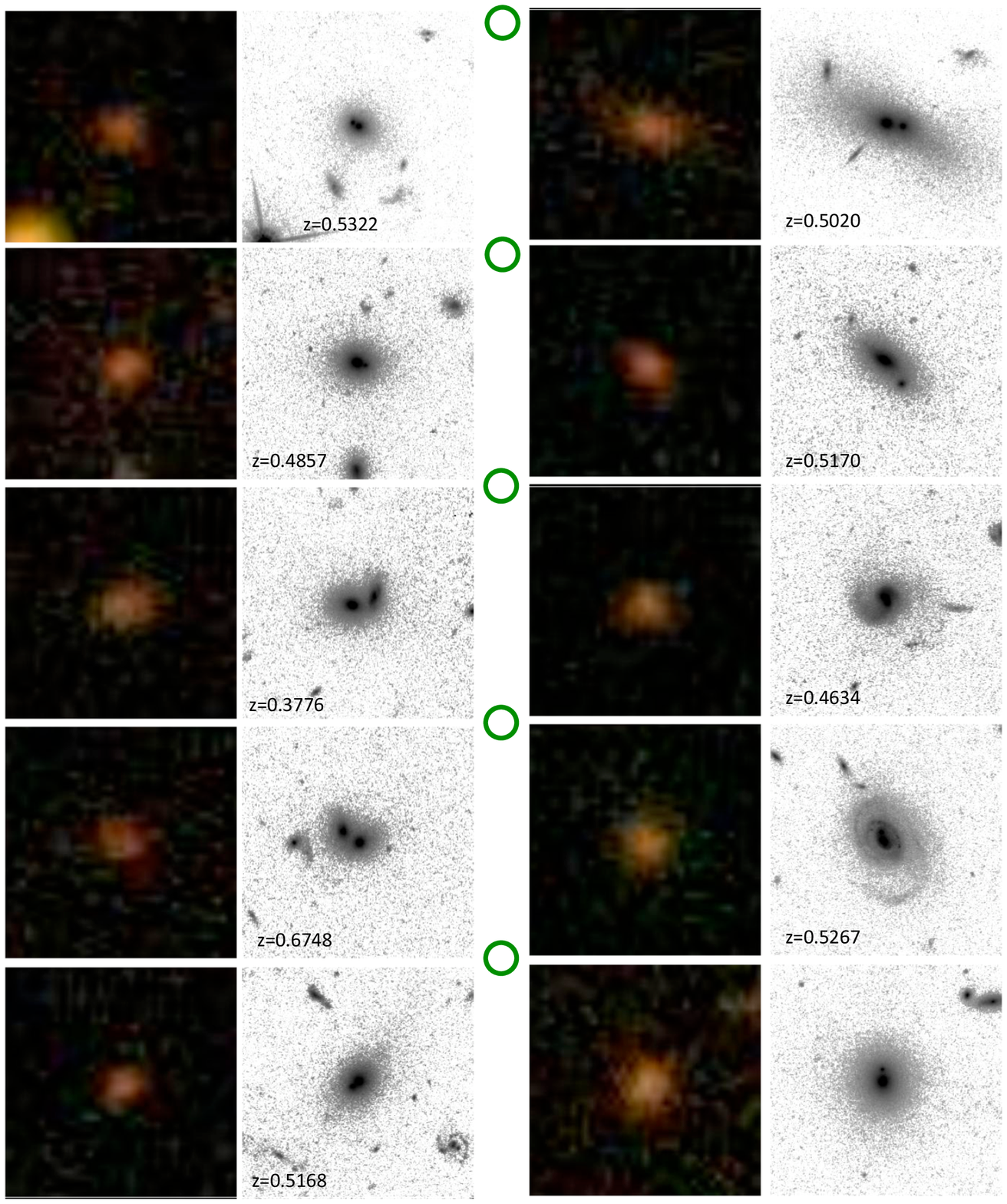}
\caption{Similar to Figures 2 and 3, this shows examples of SDSS colour composite $gri$ images, and ACS I-band images (black on white) for 10 of the BOSS CMASS galaxies which we find to have multiple components when observed at the higher HST resolution. All images are 15\arcsec square, and the BOSS 2\arcsec diameter fibre is illustrated by the green circles. The multiple components are not always very clear at the contrast level shown here, but may be inspected in more details (and in the colour composite versions) at {\tt http://icg.port.ac.uk/$\sim$mastersk/BOSSmorphologies/}. The redshifts of these galaxies from BOSS are indicated, (no redshift is known for the bottom right system). 
\label{multiples}}
\end{figure*}

We observe that 23$\pm4\%$ of the BOSS early--type galaxies are multiple systems (i.e. have multiple bright spots) in the HST data, yet are unresolved in the SDSS imaging (we find $23\pm$5\% in both CMASS and LOZ early-type galaxy subsamples). We show example images of 10 of the CMASS multiple systems in Figure \ref{multiples}. Since these objects are blended in the SDSS images (which have a seeing disc of typical size 1.1\arcsec), the typical angular separation is around  1.1\arcsec. 

Some of these multiple {\it early-type} galaxies may be ongoing (``dry") mergers \citep[e.g.][]{vd99,N06} as the colour and morphology of the sub-components are consistently red and elliptical in shape. However, only a small fraction of them (5/40) show clear evidence for interacting (having a disturbed morphology) whereas most show no evidence for tidal features and therefore could be projection effects. On the other hand, \citet{B06} argue that even in a single--orbit HST image, the morphological signature of interactions will be weak, so we may not expect many of our multiple systems to appear disturbed even if they are truly interacting. In addition to the 40 close pairs, we find 21 of the early-type galaxies have a disturbed morphology, and a further 6 have clear dust lanes (which are often associated with former mergers, e.g. Kaviraj et al. 2010). Taken to the extreme, these findings may imply that up to 67 of the 178 early-type galaxies (38$\pm5$\%) may show evidence for ongoing or recent interactions.

The multiple fraction is lower for late-type BOSS galaxies than for the early-types. In the CMASS (higher redshift) we find $15\pm7$\% late-type multiples (or 5/34 objects), while in the LOZ (lower redshift) sample there are only two multiples out of 25 late-type galaxy systems. Most of these late-type multiples (5/6 in total - note that one system satisfies both the CMASS and LOZ selection) also show a disturbed morphology; in addition we find five more late-type galaxies with evidence of a disturbed morphology. 

We first investigate if these multiple systems are projection effects. This is can be estimated using the published angular correction function, w($\theta$), of the COSMOS galaxies by McCracken et al. (2007). They find w($\theta$)$\simeq1$ on scales of an arcsecond, the typical seeing of the BOSS data, for the brightest magnitude slice they considered ($21<i_{AB}<22$). This implies there are approximately twice as many observed pairs of galaxies in the COSMOS data at these small angular scales than expected from random associations, and there appears to be a steepening of the COSMOS correlation functions on these scales (see Figure 1 of McCracken et al. 2007). This estimate crudely agrees with the available $z$COSMOS spectroscopy (Lilly et al. 2007) which provides redshifts for both components of two of our multiple systems (both early-types galaxies). One of these two multiple systems is a projection effect, while the other is a real association. A more detailed analysis of possible projection effects could be obtained from studying the angular clustering of COSMOS galaxies, with approximately the same colors and magnitudes as our BOSS galaxies, and/or studying the BOSS spectra for multiple redshifts (due to overlapping spectra from the components). Such analyses may become important in the future to fully understand all the systematic biases in the BOSS clustering measurements. For example, Le Fevre et al. (2000)  describe such a bias introduced into ground-based flux-limited samples (that lack the resolution to separate close pairs) which can cause a Òmerger-induced luminosity enhancementsÓ, i.e., the additional flux from a close companion can cause dimmer (or less massive) galaxies than desired to be added to the sample.

Separations of the components in the HST images are measured to go up to 2.0\arcsec, however this is for an object with worse than average SDSS seeing of 1.4\arcsec. A more typical separation for the majority of well separated components is 1.2\arcsec. For the $\geq 50\%$ of multiples that are physically associated, we use this to derive a typical separation of $7.3$ kpc assuming the mean CMASS redshift of $z=0.5$. This value of projected separation suggests that the pairs lie within a single dark matter halo which would have a relatively short timescale for merging (Rix \& White 1989), perhaps less than half a gigayear \citep{B06}. These physically-associated multiple early-type BOSS galaxies will have merged into a single galaxy by $z=0$ since $z=0.5$ represents a lookback time of five gigayears.

Visual inspection suggests that up to half of the multiples could be major mergers (defined here as the components having flux ratios within the range of 1:4 of each other, and both components having similar optical colours). Unfortunately, the COSMOS pipeline photometry is likely to be unreliable for such close galaxy pairs which are not easily deblended, and a more quantitative analysis of the photometry of these multiples, to accurately measure their flux ratios and estimate their mass ratios and photometric redshifts, is beyond the scope of this paper. If approximately half of our multiples are indeed major mergers (and at least half are physically associated), this might appear to contradict the low rate of major mergers seen previously for massive galaxies \citep[e.g.][]{LeF00,Ma05,J09,R09,Bu09}. However, we note that it has been argued that morphologically selected mergers (particularly in optical imaging) tend to be preferentially minor mergers \citep{B04,B10,L10}, and also the range of major merger rates published for different samples with different colour and redshift is rather accomodating  (see e.g. Table 4 of Tojeiro \& Percival (2010) which lists merger rates for massive galaxies from various samples of between 0.6-10\% per Gyr\footnote{The full BOSS redshift range spans almost 4 Gyr}, and also Section 7 of Wake et al. (2008) which discusses the difficulty of performing such comparisons). A more careful analysis will be required to properly derive the major merger rate but the relatively high fraction of unresolved multiples remains an interesting observation from our COSMOS sample of BOSS galaxies.


\subsection{Comparison with ZEST Types}

The Zurich Estimator of Structural Type (ZEST, Scarlata et al. 2007) uses a principal component analysis of five non-parametric measurements of a galaxy structure (asymmetry, concentration, a measure of uniformity, the second order moment of bright pixels and the ellipticity) to assign basic morphological types to COSMOS galaxies. ZEST classifications are available for 231 of our BOSS targets (unresolved sources and targets near the edges of COSMOS images are not included). In the ZEST catalogue, types are parameterised by a morphological code (T=1 for early-type galaxies, T=2 for disc galaxies, and T=3 for irregulars), while for disc galaxies a ``bulginess" parameter is also given, ranging from B=1 for bulge dominated galaxies to B=4 for a pure disc galaxy. 
 
\begin{figure*}
\includegraphics[width=15cm]{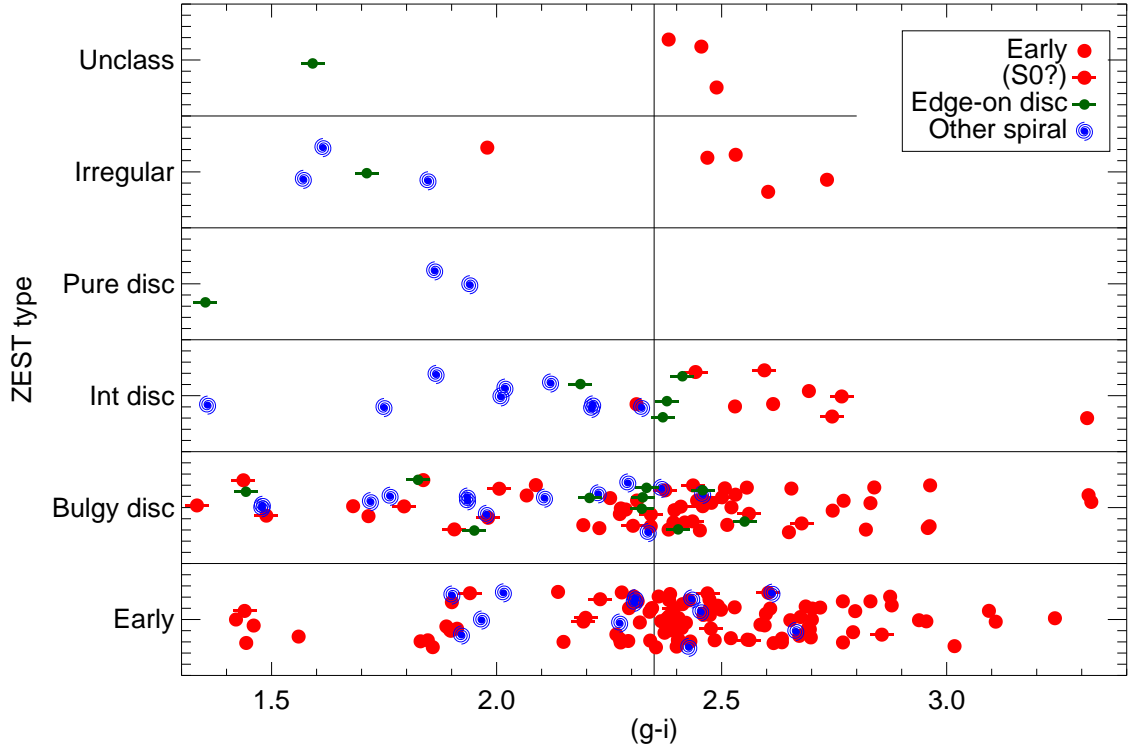}
\caption{A comparison between our visual morphological types of BOSS galaxies and the automatically generated morphology in the ZEST catalogue (Scarlata et al. 2007).  We split the ZEST types into ``early" (T=1), ``bulgy disc" (T=2, and B=1), ``intermediate discs" (T=2, and B=2 or 3), ``pure disc" (T=2 and B=4), ``irregular" (T=3) and ``unclassified" (T=9). We have artificially scattered the data points in the y-direction within each category. Along the $x$-direction we plot the $(g-i)$ colour which we showed in Section 3.1 can be used to divide the galaxy population. The vertical solid line shows our $(g-i)=2.35$ divider. The median error in this colour combination for the BOSS-COSMOS sample is $\sigma_{(g-i)}=0.1$. 
\label{typecompare}}
\end{figure*}

 Figure \ref{typecompare} shows a comparison of the ZEST types with our visual classifications. On the x-axis is the SDSS $(g-i)$ colour, which we have shown to correlate with our visual types. We split the ZEST types along the y-axis into bins representing ``early" (T=1), ``bulgy disc" (T=2 and B=1), ``intermediate disc" (T=2 and B=2 or 3), ``pure disc" (T=2 and B=4), ``irregular" (T=3) and finally ``unclassified" (T=9). We find a good overall agreement between ZEST and our visual classifications. For example, amongst the ZEST identified early-type galaxies in our sample, we identify 90\% (103/115) with the same morphology. Amongst the ZEST intermediate and pure disc galaxies, we identify 77\% (17/22) as having a disc (late-types or S0s). 
 Overall, we conclude that our visual identifications are in excellent agreement with those from the ZEST catalogue, i.e., up to 90\% of the time the two classifications agree, with the most uncertain galaxies being the early-type disc galaxies (lenticulars). 

\subsection{Shape and morphology: the Axial Ratio\label{cut}}

 We measure the shape of BOSS target galaxies in SDSS imaging by using the axial ratio, $a/b$,  from the exponential model fit to the $i$-band data. We chose this rather than the de Vaucouleurs model fit, as only in disc galaxies should this quantity differ significantly from unity (round) when such objects are viewed nearly edge-on. We have checked the difference between the exponential and de Vaucouleurs fit measurements of the axial ratio and find it to be small ($\Delta \log(a/b)=-0.01\pm$0.21) with no obvious trend with visual morphology. We have also checked this value against the axial ratios from the {\tt GIM2D} fits of ZEST and again find the difference to be small ($\Delta \log(a/b)=0.01\pm0.09$). 
 
\begin{figure}
\includegraphics[width=84mm]{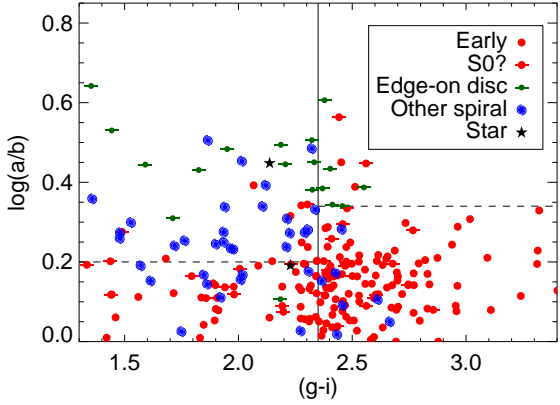}
\caption{The $(g-i)$ colour versus axial ratio (from the $i$-band exponential fit) for the 240 BOSS targets with COSMOS imaging. The vertical line illustrates our suggested cut at $(g-i)=2.35$, while the horizontal dashed lines are at $\log(a/b)=0.2$ and $\log(a/b)=0.34$, which remove most of the late-type BOSS galaxies and edge--on spirals.  
 \label{shapecolour}}
\end{figure}

 In Figure \ref{shapecolour}, we show all 240 BOSS objects in the COSMOS field as a function of visual morphology, colour and axial ratio (expressed as $\log(a/b)$).  We find that almost all our BOSS targets in the COSMOS field have an axial ratio of $a/b < 5$, or $\log(a/b)<0.7$. The only objects with highly elongated shapes in our field (i.e., $a/b>5$) are the two objects which turn out to be composed of pairs of points sources and which have $a/b=20$ (not shown in Figure). This fact could be used to remove such binary point source contamination from the remaining BOSS galaxy targets.
 
  For targets which are visually confirmed galaxies, the axial ratios behave as expected, with early-type galaxies being preferentially rounder and the most elongated galaxies being inclined spiral galaxies (see e.g. Alam \& Ryden (2002), Padilla \& Strauss (2008) for a discussion of the shape, colour and morphologies of the local galaxy population). However, we point out that the rounding effects of seeing (e.g. as described in Masters et al. 2003, and very obvious in the SDSS images shown in Figures 1 \& 2) make even the most inclined BOSS late-types relatively round objects, so only qualitative comparison with local studies should be attempted.

The distribution of the axial ratios of our early-type galaxies is consistent with the
findings of morphological studies performed with local galaxies.
Early-types show a dichotomy with bright ellipticals that are rounder
while low-luminosity ellipticals and S0 can be flatter if viewed edge-on (Kormendy \& Bender 1996).
This classification has been confirmed and extended with the SAURON sample (de Zeeuw et al. 2002) introducing
a new parameter based on the angular momentum of the galaxy that separates early-types
in slow and fast rotators (Emsellem et al. 2007 and Cappellari et al. 2007).
Slow rotators tend to be quite round while the fast rotators can be more
flattened (again depending on viewing angle). This appears similar to Figure 7, in which the bulk of early-types
(mainly bright E, i.e. slow rotators) have $\log(a/b)<0.2$, but there are some with
$\log(a/b)>0.2$ that could correspond to the E-S0 SAURON fast-rotators.
Recent work on the morphology of nearby galaxies (Cappellari et
al. 2011a,b, Emsellem et al. 2011) has confirmed the differences in the
axial ratios described above using the much wider ATLAS3D sample
(Cappellari et al. 2011a).

All the inclined spiral galaxies in our BOSS sample could be removed by requiring the SDSS axial ratio to be $\log(a/b)<0.34$. The only exception is a galaxy near a bright star. This axial-ratio cut is particularly useful in the LOZ sample where, as illustrated in Figure \ref{colour}, the colour cut alone leaves a substantial number of edge-on late-type galaxies in the sample. 

\subsection{Redshift, Luminosity and Morphology}
\begin{figure}
\includegraphics[width=84mm]{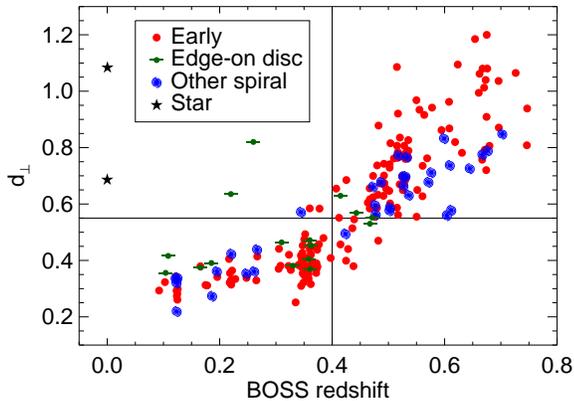}
\caption{Redshifts of targets plotted against $d_{\perp}=(r-i)-((g-r)/8)$ for the 224 targets with reliable redshifts in the COSMOS field. The horizontal line is 
$d_{\perp}=0.55$ which is the selection used for the CMASS subset. The vertical line shows $z=0.4$. 
 \label{redshiftmorph}}
\end{figure}

In Figure \ref{redshiftmorph} we show the BOSS targets with reliable redshifts in the COSMOS field versus a combination of $gri$ photometry,  $d_\perp=(r-i)-(g-r)/8.0$, which is used in the target selection (see Section 2.3). Symbols indicate the visual morphology. By design, this $gri$ colour combination tracks the redshift of a passively--evolved stellar population and a cut of $d_\perp>0.55$ selects the higher redshift CMASS subset in BOSS (shown by a horizontal line in Figure \ref{redshiftmorph}). It has already been demonstrated that this cut accurately isolates $z>0.4$ galaxies in BOSS (White et al. 2011). We show here that this is true regardless of the visual morphology of the BOSS galaxies. We find that in this higher redshift subset, only two significantly lower redshift galaxy interlopers are present (or $0.8\pm$0.6\% of the resulting sample), both of which are inclined spiral galaxies which we suggest may be dust reddened into the selection of $d_\perp>0.55$. It is interesting to notice that in Figure 1 of White et al. (2011), which shows the redshift distribution of the first semester of CMASS galaxies, there is a small bump present at $z\sim0.2$ which we now suggest is probably explained by these inclined spiral interlopers into the CMASS colour selection. 

 We also point out in Figure \ref{redshiftmorph} the clear over-density at $z=0.35$ in the COSMOS field, which has previously been noticed \citep[e.g.][]{cosmosdensity} and which clearly illustrates the well known morphology-density relation. 
  

\begin{figure}
\includegraphics[width=84mm]{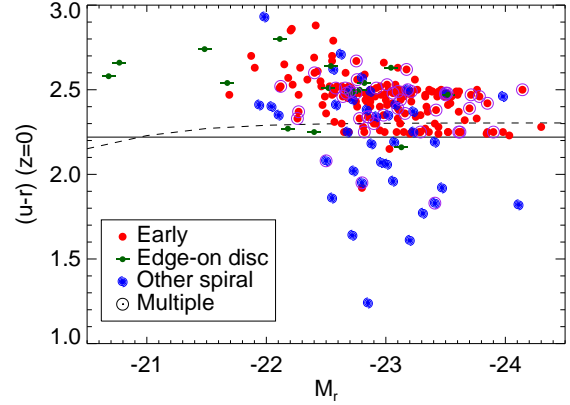}
\caption{\klm{Colour--magnitude (CM) diagram showing $(u-r)$ and $M_r$  k-corrected to $z=0$ for  BOSS galaxies in the COSMOS field with redshifts.  The horizontal line shows $(u-r)=2.22$ derived as an optimal separator for the $z=0$ galaxy population by Strateva et al. (2001), while the dashed line is the similar magnitude dependent separator derived by Baldry et al. (2004). Since BOSS targets only massive, luminous galaxies, they lie at the luminous end of the local CM where the main locus of the red sequence and blue cloud lie close together. }
\label{cm}}
\end{figure}

In Figure \ref{cm}, we plot the rest-frame colour $(u-r)$ versus the observed-frame absolute magnitude ($M_r$) for the COSMOS BOSS galaxies with redshifts. The k-corrections are derived using fits to the mixed model of Maraston et al. (2006) where the galaxies have $(g-i)_{\rm obs}<2.35$ and to the LRG model of Maraston et al. (2009) otherwise. We show for comparison the colour dividers derived by Strateva et al. (2001; solid line) and Baldry et al. (2004; dashed line) for the local galaxy population. Most BOSS targets have (rest frame) $r$-band absolute magnitudes above $M_r<-21.5$ and thus represent the high luminosity end of the colour--magnitude relationship, where the main loci of the red sequence and blue cloud move close together in colour space. Never the less we find that the BOSS late-type galaxies lie preferentially in the blue cloud.  We point out that no internal extinction corrections have been applied for the disc galaxies - so the dust reddening may be significant (up to half a magnitude, Masters et al. 2010a) for the edge-on discs (green symbols). We also indicate in Figure \ref{cm} if the galaxies are unresolved multiples, and therefore possibly merger events. Such objects have no obvious offset in this colour--magnitude diagram, indicating there is no evidence for merger-induced star formation. 

\section{Catalogued Sizes of BOSS galaxies}

 The sizes of massive galaxies at high redshifts have been a topic of significant interest in recent years (e.g. Trujillo et al. 2006, Cimatti et al. 2008, Mancini et al. 2009, Shankar \& Bernardi 2009, van der Wel et al. 2009, Maltby et al. 2010), providing new constraints on hierarchical galaxy evolution models \citep[e.g.][]{A07,shankar2010,shankar2011}. Many observations of massive galaxies at $z>1$ suggest they are significantly more compact than their counterparts at $z=0$ (e.g. Daddi et al. 2005; Trujillo et al. 2006; Longhetti et al. 2007; Cimatti et al. 2008; van Dokkum et al. 2008, van der Wel et al. 2008). This size evolution is in qualitative agreement with the models, but indicates stronger evolution than is usually predicted (although Mancini et al. 2009 have argued the observed  size evolution may be over estimated due to the lower $S/N$ of high redshift observations). Theses studies have largely focussed on comparing high redshift ($z>1$) galaxies with the local galaxy population. The CMASS sample of BOSS, representing massive luminous galaxies at an intermediate redshift range of $0.4<z<0.7$ will therefore be an important sample to link the two populations, and further test such models. Accurate size measurements for all BOSS galaxies, but particularly the CMASS subset are therefore of significant interest, important not only for this test, but also for measuring accurate dynamical masses (see e.g. Beifiori et al. in prep.).
  
 The median (seeing-corrected) effective radius of our sample of BOSS galaxies is 1.5\arcsec ~(1.2\arcsec for CMASS and 1.9\arcsec for LOZ), taken from the de Vaucouleurs fit to either the SDSS $r$- or $i$-band images. This is close to the median seeing of the SDSS-III DR8 imaging (Section 2.2). In this section, we will compare the SDSS pipeline size measurements for CMASS galaxies to those measured from COSMOS imaging provided by ZEST \citep{ZEST}. There are 129 CMASS targets in the COSMOS field, of which 123 have sizes reported in ZEST (the six with no reported size comprise the four shown to be made up of point sources in the HST  imaging, and two galaxies which are near the edges of COSMOS HST images). 

ZEST provides Sersic fits to the HST galaxy profiles using the {\tt GIM2D} software \citep{sargent07}{\footnote{We use the parameter {\tt R\_GIM2D} from the ZEST fits, which we point out is the effective radius from the version of the Sersic profile in which this radius is defined to be identical to the half-light radius, \ie~ $I(r)=I(0) \exp(-b_n (r/r_{e})^{1/n})$, where the constant $b_n$ is a function of the shape parameter, $n$, chosen such than $r_e=r_{1/2}$, and is well approximated by $b_n=2n-0.324$ for $1<n<15$ \citep{TGC01}}, while the effective radii from the SDSS pipeline analysis are based on a de Vaucouleur fit, i.e., a $n=4$ Sersic profile. Here we present a simple comparison between the SDSS pipeline (de Vaucouleurs model) and ZEST (Sersic model) effective radii. While it is clear that if the true galaxy profile is a Sersic profile, (\ie~ with arbitrary $n$) then forcing it to an $n=4$ profile while fitting  will affect the estimate of the effective radius (e.g. see  Figure 4 of \citet{TGC01} which shows the analytic description of how this approach may change the effective radius when using ideal profiles with no seeing, or random noise), we argue that this effect is not likely to be the dominant bias on the effective radii of CMASS galaxies measured by SDSS. In this case, the seeing is comparable in size to the effective radius, so the profile shape fit is actually dominated by the point--spread function, and furthermore is constrained by a small number (5 or 6) data points. To illustrate this effect, we show in Figure \ref{profiles} four example profiles of BOSS galaxies compared to a Gaussian point--spread function (solid line; we note that in SDSS the seeing disc is actually fit as a double Gaussian which would extend to even larger radii than the single Gaussian shown as a guide here), a de Vaucouleurs profile (dashed line) and a Sersic profile (dotted line).  


The effect of seeing on the SDSS profiles is so significant, that it is almost impossible to distinguish between the Sersic and de Vaucouleurs fits. As will be shown in Beifiori et al. (in prep.) it is not possible to fit Sersic profiles to the SDSS imaging data of galaxies at the redshifts of BOSS galaxies as there is not enough information to constrain both $n$ and the effective radius, and in addition the strong correlation between $n$ and $r_e$ means that errors in the value of $n$ will result in changes in the inferred value of $r_e$. 

We show the relative difference between the SDSS pipeline and ZEST sizes as a function of the ZEST size in Fig. \ref{size1}. We see no trends of the difference with the ZEST Sersic fit $n$, and the difference in effective radius between the two pipelines does not follow the theoretical expectation for best fit de Vaucouleurs values to a Sersic profile. We argue that this is further evidence that the dominant bias is not due to fitting an $r^{1/4}$ model in the SDSS images. The median relative difference between SDSS pipeline effective radii of CMASS targets (in $i$-band) and those from ZEST fits is +43$\pm90$\% (i.e. SDSS sizes are on average 43\% larger) but the relative measurement difference ranges from -50\% to +500\%. In absolute value this full range corresponds to -0.8--2.7\arcsec, and the median absolute difference is $+0.31\pm0.67$\arcsec.

\begin{figure}
\includegraphics[width=84mm]{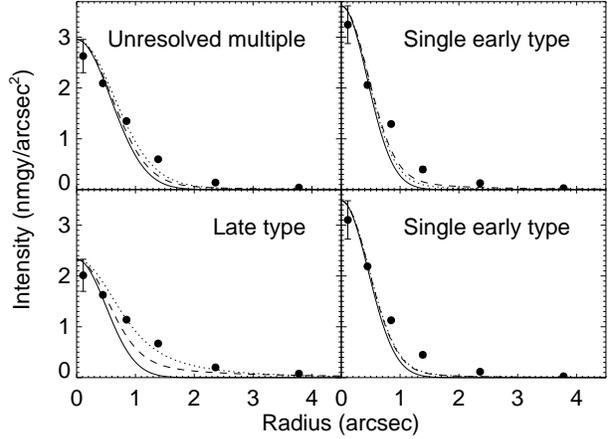}
\caption{Example SDSS $i$-band profiles of four CMASS galaxies. Shown are a Gaussian PSF (solid line), the SDSS de Vaucouleurs fit (dashed line) and the ZEST Sersic fit (dotted line). The latter two models are convolved with a single Gaussian PSF.  There is a y offset applied to all three profiles such that they set to be equal to the maximum of the central error range at $r=0$. 
\label{profiles}}
\end{figure}

\begin{figure}
\includegraphics[width=84mm]{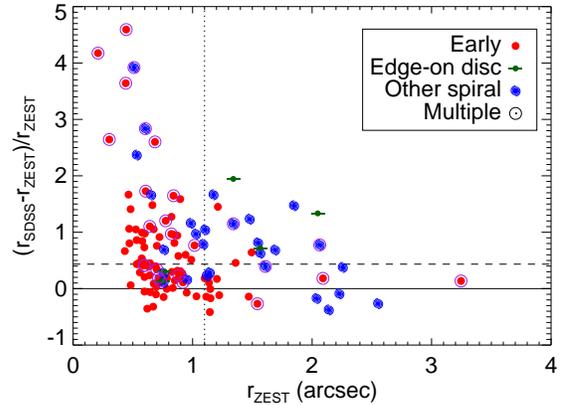}
\caption{Relative difference between the SDSS pipeline effective radius and the effective radius reported in ZEST as a function of the ZEST size. The median difference of 43\% is shown with the dashed line. Visual morphologies are indicated by the symbols as per the legend. Where two symbols overlap the galaxy is found in both categories (e.g. multiple early types). 
 \label{size1}}
\end{figure}

 As might be expected, the most catastrophic errors in the SDSS sizes occur when the BOSS target is an unresolved multiple in the SDSS imaging. In these cases the ZEST size measurements is of a single component, making it much smaller than the SDSS size which represents the separation of the multiple system. Almost all objects with relative differences greater than +200\% are unresolved multiples which have ZEST sizes $\simeq 0.5$\arcsec, but SDSS sizes $>1.5$\arcsec ~(see Figure \ref{size1}). Such objects cannot be removed using any cuts described above with SDSS data, and we have yet to find any alternative criterion based on SDSS data alone which can remove them.

There is no significant improvement in the median relative difference (i.e.. it remains at $+40\%$) by removing blue (i.e.. $(g-i)<2.35)$) or highly elongated ($\log(a/b)>0.34$) galaxies from CMASS.  We also see no evidence for trends of the size difference with colour, morphology (either visual, or using the index of the ZEST sersic fit, as discussed above, and also no trends with redshift. 
 There are also no trends of the difference with the size of the SDSS seeing disc (which varies from 0.95 to 1.4\arcsec) in the COSMOS area. Also there are no correlations with the reported likelihood of the SDSS de Vaucouleurs fit, or the reported error on the SDSS effective radius. 
 
  One caveat here is that the BOSS target galaxies are very large galaxies for COSMOS. It has been shown that Sersic profile fits will underestimate effective radii in cases where the sky is marked too close to the galaxies \citep[as first noted by][]{TGC01}. It seems feasible that this could be a problem especially for the larger galaxies in COSMOS.  We will present a much more comprehensive comparison of various size measurements of BOSS target galaxies considering issues like this further
in a future paper (Beifiori et al. in prep.).

\section{Conclusions}

We present a study of the visual morphology and measured size of galaxies from the Baryon Oscillation Spectroscopic Survey (BOSS). We focus on a sample of BOSS galaxies that overlap with the COSMic Origins Survey (COSMOS), which possess high resolution, deep HST imaging data. Our major conclusions are:

\begin{itemize}

\item
We find that 76$\pm3$\% of BOSS galaxies in both the ``CMASS" and ``LOZ" samples have an early-type visual morphology and the red colours typical of passive galaxies at the targeted redshift. This is in agreement with the modeling of the survey target selection. The remainder of BOSS galaxies have a late--type morphology. 

\item 
We show that a simple colour cut of $(g-i)=2.35$ serves to remove most of the late-type galaxies in both the CMASS and LOZ subsamples of BOSS galaxies, leaving each sample with $\ge90$\% early-type galaxies. We explain why this colour cut works using models, \klm{and point out that it is similar to the rest frame colour cut of $(u-r)=2.22$ derived for local galaxies (Strateva et al. 2001)}. Likewise, a simple cut of $\log(a/b)<0.34$ on the axial ratio of BOSS galaxies would remove binary star systems and edge-on disc galaxies. Together, these cuts would produce as homogeneous a sample of BOSS galaxies as presently possible, which may be required for some studies of galaxy evolution and cosmology.   

\item We find more edge-on late-type galaxies in the LOZ subset than expected from random orientations on the sky, suggesting that this selection contains a subsample of inclined spirals preferentially making the sample selection. This problem may become a major systematic effect for detailed studies of the BOSS sample, e.g., photometric redshift estimations (see Yip et al. 2011; Ross et al. 2011). The colours of these inclined spirals are however still not as red as the reddest early--type galaxies in our sample, indicating that the colours of BOSS galaxies are dominated by their redshift and stellar populations.  

\item We find intrinsically red spiral galaxies (\ie ~not highly inclined), which maybe analogous to the ``passive red spirals" found recently in both the local Galaxy Zoo data-set \citep{M10} and in the COSMOS field \citep{B10}. These objects make up $10\pm3\%$ of the galaxies in the sample above $(g-r)=2.35$. 

\item 
For the early--type BOSS galaxies, we find a significant fraction ($23\pm4$\%) are unresolved multiples in the SDSS imaging. We estimate that at least 50\% of these multiples are likely real associations and not projection effects. For those which are physically associated, the components of these multiple system are relatively close ($\sim7$ kpc at the median redshift of BOSS), 
and we conclude that many are likely on--going ``dry mergers" that reside in the same dark matter halo and will merge completely on a short timescale \citep{B06}. 

\item
 The standard SDSS pipeline measurements of sizes for BOSS galaxies have a median size of 1.5\arcsec (1.2\arcsec ~for CMASS and 1.9\arcsec for LOZ), meaning that the bulk of BOSS galaxies are barely resolved in the SDSS imaging. SDSS measurements of the seeing--corrected effective radius of CMASS galaxies are on average  0.3\arcsec (40\%) larger than similar measurements made from the HST imaging (using the ZEST catalogue of \citealt{ZEST}) but show no trend with colour, morphology or redshift. A correction of $-0.31\pm0.67$\arcsec can be applied to statistically correct the SDSS pipeline size measurements to make them agree with those from the HST imaging (but there will still be significant error on an individual galaxy).

\end{itemize}

In this paper, we have demonstrated the effectiveness of the SDSS-III BOSS target selection which was aimed at selecting massive, luminous galaxies regardless of their morphology or star formation history. Most such galaxies are early-type and passive galaxies, but there exists a significant fraction of massive late-type galaxies.} As shown in Figures 8 and 9, BOSS galaxies are predominantly passive, early--type, massive galaxies over the redshift range $0.1<z<0.7$. Approximately a quarter of BOSS galaxies are late-types. Depending on the science application, this subset of galaxies could represent a potential systematic uncertainty. The results of this paper will help users of the BOSS data to fine-tune their selections to obtain the sample of galaxies they require. 

\paragraph*{ACKNOWLEDGEMENTS.} 

KLM acknowledges funding from The Leverhulme Trust through a 2010 Early Career Fellowship. RT also thanks the Leverhulme trust for financial support. CM, RCN, DT, AB, EME, and AJR acknowledge STFC rolling grant ST/I001204/1 ``Survey Cosmology and Astrophysics" for support.

Funding for SDSS-III has been provided by the Alfred P. Sloan Foundation, the Participating Institutions, the US National
Science Foundation, and the U.S. Department of Energy. The SDSS-III web site is {\tt http://www.sdss3.org/}. 

SDSS-III is managed by the Astrophysical Research Consortium for the Participating Institutions of the SDSS-III Collaboration including the University of Arizona, the Brazilian Participation Group, Brookhaven National Laboratory, University of Cambridge, University of Florida, the French Participation Group, the German Participation Group, the Instituto
de Astrofisica de Canarias, the Michigan State/Notre Dame/JINA Participation Group, Johns Hopkins University,
Lawrence Berkeley National Laboratory, Max Planck Institute for Astrophysics, New Mexico State University, New
York University, the Ohio State University, the Penn State University, University of Portsmouth, Princeton University,
University of Tokyo, the University of Utah, Vanderbilt University, University of Virginia, University of Washington, and Yale University.

This research has made use of the NASA/ IPAC Infrared Science Archive, which is operated by the Jet Propulsion Laboratory, California Institute of Technology, under contract with the National Aeronautics and Space Administration. 

Some of the data presented in this paper were obtained from the Multimission Archive at the Space Telescope Science Institute (MAST). STScI is operated by the Association of Universities for Research in Astronomy, Inc., under NASA contract NAS5-26555. Support for MAST for non-HST data is provided by the NASA Office of Space Science via grant NNX09AF08G and by other grants and contracts.

This publication made use of observations made with the NASA/ESA Hubble Space Telescope, and obtained from the Hubble Legacy Archive, which is a collaboration between the Space Telescope Science Institute (STScI/NASA), the Space Telescope European Coordinating Facility (ST-ECF/ESA) and the Canadian Astronomy Data Centre (CADC/NRC/CSA).

\appendix
\section{Data Table}
We provide here a sample of the basic data available on our sample of 240 BOSS galaxies. We list the equatorial coordinates of the galaxy (RA and Dec in J2000), $(g-i)$ colour, redshift and (where available), the BOSS subsample for the galaxy (see Section 2.3) and the visual morphological type we assign (see Section 3). A page of the table (in which galaxies are presented in order of increasing RA). is shown below to illustrate the format. The full table is available electronically. \\

\begin{table*}
\caption{The BOSS-COSMOS cross match sample (The full table is available in the online version of the paper.)}
\label{redtable}
\begin{tabular}{cccclll}
\hline
RA (J2000 deg) & Dec (J2000 deg) & $(g-i)$ & Redshift & Redshift source & BOSS Subsample & Visual type \\
\hline
 149.45591 &    2.56332 &   2.69 &  0.3732 & BOSS &  LOZ &  Early \\
 149.46232 &    2.00290 &   2.23 &  0.3776 & BOSS & CMASS &   Early disturbed multiple \\
 149.46780 &    2.67181 &   2.42 &  0.4284 & BOSS &  LOZ &  Early \\
 149.47437 &    1.97636 &   2.23 &  0.0004 & BOSS & CMASS &   Star \\
 149.47484 &    1.95824 &   1.59 &  0.1084 & SDSS2 &  LOZ &  Late edge \\
 149.47913 &    2.56491 &   2.43 &  0.3717 & SDSS2 &  LOZ &  Early \\
 149.48440 &    2.39573 &   2.68 &  0.4819 & BOSS &  LOZ &  Early \\
 149.48441 &    2.43524 &   2.72 &  0.4800 & BOSS & CMASS &   Early \\
 149.48569 &    1.81567 &   2.70 &  0.5301 & BOSS & CMASS &   Early \\
 149.49280 &    2.15313 &   1.49 &  0.1254 & SDSS2 &  LOZ &  Early S0 \\
 149.49475 &    1.70034 &   1.94 &  0.5714 & BOSS & CMASS &   Late \\
 149.49892 &    1.77765 &   1.76 &  0.5030 & BOSS & CMASS &   Late bar disturbed \\
 149.50357 &    2.02399 &   2.53 &  0.6739 & BOSS & CMASS &   Early multiple \\
 149.50510 &    1.93552 &   2.82 &  0.5356 & BOSS & CMASS &   Early multiple \\
 149.52362 &    1.75892 &   2.84 &  0.5688 & BOSS & CMASS &   Early \\
 149.52782 &    1.93026 &   2.70 &  0.5354 & BOSS & CMASS &   Early multiple \\
 149.53110 &    2.44029 &   2.60 &  0.4960 & BOSS & CMASS &   Early disturbed \\
 149.53903 &    1.72237 &   2.56 &  0.5016 & BOSS & CMASS &   Early \\
 149.55600 &    2.52001 &   2.53 &  0.4857 & BOSS & CMASS &   Early multiple \\
 149.57499 &    2.40669 &   2.43 &  0.4825 & BOSS & CMASS &   Early \\
 149.58221 &    2.83479 &   2.35 &  0.3485 & BOSS &  LOZ &  Early multiple \\
 149.60008 &    2.82114 &   2.40 &  0.3446 & SDSS2 &  LOZ &  Early \\
 149.60469 &    2.45263 &   2.01 &  0.6768 & BOSS & CMASS &   Late \\
 149.61937 &    1.73545 &   2.96 &  0.5947 & BOSS & CMASS &   Early \\
 149.61972 &    2.25936 &   2.52 &  0.3602 & BOSS & CMASS &   Early \\
 149.63300 &    2.79217 &   2.29 &  0.4123 & BOSS & CMASS &   Early \\
 149.63728 &    2.60176 &   3.02 &  0.5499 & BOSS & CMASS &   Early \\
 149.63763 &    1.87079 &   2.37 & - & No redshift & CMASS &   Late bar \\
 149.64264 &    2.47120 &   2.32 &  0.5264 & BOSS & CMASS &   Late \\
 149.64834 &    2.81872 &   2.43 &  0.3446 & SDSS2 & CMASS and LOZ &  Late bar disturbed multiple \\
 149.65295 &    1.65853 &   2.55 &  0.4672 & BOSS &  LOZ &  Late edge \\
 149.65426 &    2.85348 &   2.02 &  0.7026 & BOSS & CMASS &   Late disturbed \\
 149.66390 &    1.62584 &   2.38 &  0.4668 & BOSS & CMASS &   Early \\
 149.68306 &    1.88888 &   2.53 &  0.5322 & BOSS & CMASS &   Early multiple \\
 149.68626 &    2.33281 &   2.39 &  0.5515 & BOSS & CMASS &   Early \\
 149.68789 &    2.65381 &   2.79 & - & No redshift & CMASS &   Early \\
 149.68793 &    2.47519 &   2.44 & - & No redshift & CMASS &   Early bar S0 disturbed \\
 149.68883 &    2.49744 &   2.34 &  0.4873 & BOSS & CMASS &   Late bar \\
 149.69039 &    1.61994 &   2.77 &  0.4815 & BOSS & CMASS &   Early \\
 149.69557 &    2.40576 &   4.04 &  0.6757 & BOSS & CMASS &   Early \\
 149.70085 &    2.66303 &   2.67 &  0.4921 & BOSS & CMASS &   Early \\
 149.70565 &    2.50560 &   2.70 &  0.4913 & BOSS & CMASS &   Early \\
 149.71041 &    2.12273 &   2.56 &  0.4796 & BOSS & CMASS &   Early S0 \\
 149.71245 &    2.54465 &   2.63 &  0.4766 & BOSS & CMASS &   Early \\
 149.71374 &    2.01956 &   2.29 &  0.6446 & BOSS & CMASS &   Late disturbed multiple \\
 149.71858 &    1.63799 &   1.97 &  0.6671 & BOSS & CMASS &   Late disturbed \\
 149.72159 &    1.73891 &   2.80 &  0.6606 & BOSS & CMASS &   Early disturbed \\
 149.72647 &    2.41945 &   1.44 &  0.1248 & SDSS2 &  LOZ &  Early S0 \\
 149.72909 &    1.91729 &   2.30 &  0.3622 & BOSS &  LOZ &  Early \\
 149.73561 &    2.10550 &   2.37 &  0.5164 & BOSS & CMASS &   Early \\
 149.73964 &    1.79295 &   1.86 & - & No redshift & CMASS &   Late bar \\
 149.74285 &    2.17959 &   3.40 &  0.6308 & BOSS & CMASS &   Early multiple \\
 149.75634 &    2.79470 &   2.73 &  0.4930 & SDSS2 & CMASS &   Early multiple \\
 149.76710 &    1.96770 &   2.61 &  0.5281 & BOSS & CMASS &   Early \\
 149.77077 &    2.78284 &   2.46 &  0.5099 & BOSS & CMASS &   Early \\
 149.77704 &    2.75691 &   2.29 &  0.3456 & BOSS &  LOZ &  Early \\
 149.77779 &    2.26297 &   2.02 & q 0.4730 & BOSS & CMASS &   Late multiple \\
 149.78498 &    2.53039 &   2.56 & - & No redshift & CMASS &   Early S0 \\
 149.79785 &    1.70203 &   2.34 &  0.3503 & BOSS &  LOZ &  Early S0 disturbed \\
 149.79810 &    1.80106 &   2.96 &  0.6621 & BOSS & CMASS &   Early \\
 149.79882 &    1.97809 &   2.47 &  0.5257 & BOSS & CMASS &   Early S0 disturbed \\
 .. & .. & .. &..  &.. &.. \\
      \hline
\end{tabular}
\end{table*}


\begin{thebibliography}{}
\bibitem[Aihara et al. (2011)]{DR8paper} Aihara, H., et al. 2011, ApJS 193, 29 
\bibitem[Alam \& Ryden(2002)]{2002ApJ...570..610A} Alam, S.~M.~K., \& Ryden, B.~S.\ 2002, \apj, 570, 610 
\bibitem[Almeida et al. (2007)]{A07} Almeida, C., Baugh. C.M., Lacey, C.G. 2007,  \mnras, 376, 1711 
\bibitem[Ball et al. (2008)]{B08} Ball N. M., Loveday J., Brunner R. J., 2008, MNRAS, 383, 907 
\bibitem[Baldry et al.(2004)]{2004ApJ...600..681B} Baldry, I.~K., Glazebrook, K., Brinkmann, J., Ivezi{\'c}, {\v Z}., Lupton, R.~H., Nichol, R.~C., \& Szalay, A.~S.\ 2004, \apj, 600, 681 
\bibitem[Bamford et al.(2009)]{B09} Bamford, S.~P., et al.\ 2009, \mnras, 393, 1324
\bibitem[Bell \& de Jong(2001)]{B01} Bell, E.~F., \& de Jong, R.~S.\ 2001, \apj, 550, 212 
\bibitem[Bell et al.(2006)]{B06} Bell, E.~F., et al.\ 2006, \apj, 640, 241 
\bibitem[Blanton et al. (2005)]{blanton2005} Blanton, M.~R., Eisenstein, D., Hogg, D.~W., Schlegel, D.~J., \& Brinkmann, J.\ 2005, \apj, 629, 143 
\bibitem[Bundy et al.(2004)]{B04} Bundy, K., Fukugita, M., Ellis, R.~S., Kodama, T., \& Conselice, C.~J.\ 2004, \apjl, 601, L123 
\bibitem[Bundy et al.(2009)]{Bu09} Bundy, K., Fukugita, M., Ellis, R.~S., Targett, T.~A., Belli, S., \& Kodama, T.\ 2009, \apj, 697, 1369 
\bibitem[Bundy et al.(2010)]{B10} Bundy, K., et al.\ 2010, \apj, 719, 1969 
\bibitem[Calzetti et al.(2000)]{2000ApJ...533..682C} Calzetti, D., Armus, L., Bohlin, R.~C., Kinney, A.~L., Koornneef, J., \& Storchi-Bergmann, T.\ 2000, \apj, 533, 682 
\bibitem[Cameron(2011)]{C11} Cameron, E.\ 2011, PASA, 28, 128  
\bibitem[Cannon et al. (2006)]{C06} Cannon, R., et al.\  2006, \mnras, 372, 425 
\bibitem[Cappellari et al.(2007)]{2007MNRAS.379..418C} Cappellari, M., et al.\ 2007, \mnras, 379, 418 
\bibitem[Cappellari et al.(2011a)]{2011MNRAS.413..813C} Cappellari, M., et  al.\ 2011a, \mnras, 413, 813 
\bibitem[Cappellari et al.(2011b)]{2011arXiv1104.3545C} Cappellari, M., et al.\ 2011b, arXiv:1104.3545 
\bibitem[Cimatti et al. (2008)]{cimatti08} Cimatti A. et al., 2008, A\&A, 482, 21
\bibitem[Daddi et al. (2005)]{daddi05} Daddi E. et al., 2005, ApJ, 626, 680 
\bibitem[de Zeeuw et al.(2002)]{2002MNRAS.329..513D} de Zeeuw, P.~T., et al.\ 2002, \mnras, 329, 513 
\bibitem[Dressler(1980)]{D80} Dressler, A.\ 1980, \apj, 236, 351 
\bibitem[Eisenstein et al. (2001)]{E01} Eisenstein, D.~J., et al.\ 2001, \aj, 122, 2267 
\bibitem[Eisenstein et al. (2011)]{E11} Eisenstein, D. ~J., et al. 2011, \aj, arXiv:1101.1529 
\bibitem[Emsellem et al.(2007)]{2007MNRAS.379..401E} Emsellem, E., et al.\ 2007, \mnras, 379, 401 
\bibitem[Emsellem et al.(2011)]{2011MNRAS.tmp..687E} Emsellem, E., et al.\ 2011, \mnras, 687 
\bibitem[Fukugita et al. (1996)]{F96} Fukugita, M., Ichikawa, T., Gunn, J.E., Doi, M., Shimasaku, K., \& Schneider, D.P. 1996, AJ, 111, 1748  
\bibitem[Griffith et al. (2011)]{G11} Griffith, R. et al.  (in prep.)
\bibitem[Gunn et al. (1998)]{G98} Gunn, J.E., et al. 1998, AJ, 116, 3040
\bibitem[Gunn et al. (2006)]{G06} Gunn, J.E., et al. 2006, AJ, 131, 2332 
\bibitem[Hogg et al.(2003)]{H03} Hogg, D.~W., et al.\ 2003, \apjl, 585, L5 
\bibitem[Ilbert et al.(2009)]{I09} Ilbert, O., et al.\ 2009, \apj, 690, 1236 
\bibitem[Jogee et al.(2009)]{J09} Jogee, S., et al.\ 2009, \apj, 697, 1971 
\bibitem[Kannappan et al.(2009)]{kap2009} Kannappan, S.~J., Guie, J.~M., \& Baker, A.~J.\ 2009, \aj, 138, 579 
\bibitem[Kauffmann et al.(1993)]{K93} Kauffmann, G., White, S.~D.~M., \& Guiderdoni, B.\ 1993, \mnras, 264, 201 
\bibitem[Kaviraj et al.(2011)]{2011MNRAS.411.2148K} Kaviraj, S., Tan, K.-M., Ellis, R.~S., \& Silk, J.\ 2011, \mnras, 411, 2148 
\bibitem[Koekemoer et al.(2007)]{K07} Koekemoer, A.~M., et al.\ 2007, \apjs, 172, 196 
\bibitem[Komatsu et al.(2011)]{komatsu11} Komatsu, E., et al.\ 2011, \apjs, 192, 18 
\bibitem[Kormendy \& Bender(1996)]{1996ApJ...464L.119K} Kormendy, J., \& Bender, R.\ 1996, \apjl, 464, L119 
\bibitem[Kova{\v c} et al.(2010)]{cosmosdensity} Kova{\v c}, K., et al.\ 2010, \apj, 708, 505  
\bibitem[Leauthaud et al.(2007)]{L07} Leauthaud, A., et al.\ 2007, \apjs, 172, 219  
\bibitem[Le F{\`e}vre et al.(2000)]{LeF00} Le F{\`e}vre, O., et al.\ 2000, \mnras, 311, 565 
\bibitem[Lilly et al.(2007)]{2007ApJS..172...70L} Lilly, S.~J., et al.\  2007, \apjs, 172, 70 
\bibitem[Lintott et al.(2008)]{L08} Lintott, C.~J., et al.\ 2008, \mnras, 389, 1179 
\bibitem[Lintott et al.(2011)]{L11} Lintott, C.~J., et al.\ 2011, \mnras, 410, 166
\bibitem[Longhetti et al. (2007)]{longhetti07} Longhetti M. et al., 2007, MNRAS, 374, 614 
\bibitem[Lotz et al.(2010)]{L10} Lotz, J.~M., Jonsson, P., Cox, T.~J., \& Primack, J.~R.\ 2010, \mnras, 404, 575 
\bibitem[Maller et al.(2009)]{Ma09} Maller, A.~H., Berlind, A.~A., Blanton, M.~R., \& Hogg, D.~W.\ 2009, \apj, 691, 394 
\bibitem[Maltby et al.(2010)]{2010MNRAS.402..282M} Maltby, D.~T., et al.\ 2010, \mnras, 402, 282 
\bibitem[Mancini et al. (2009)]{mancini09} Mancini C., Matute I., Cimatti A., Daddi E., Dickinson M., Rodighiero G., Bolzonella M., Pozzetti L., 2009, A\&A,  500, 705 size 
\bibitem[Maraston(2005)]{M05} Maraston, C.\ 2005, \mnras, 362, 799 
\bibitem[Maraston et al.(2006)]{M06} Maraston, C., Daddi, E., Renzini, A., Cimatti, A., Dickinson, M., Papovich, C., Pasquali, A., \& Pirzkal, N.\ 2006, \apj, 652, 85 
\bibitem[Maraston et al. (2009)]{M09} Maraston, C., Str{\"o}mb{\"a}ck, G., Thomas, D., Wake, D.~A., \& Nichol, R.~C.\ 2009, \mnras, 394, L107 
\bibitem[Maraston et al.(2010)]{CM10} Maraston, C., Pforr,  J., Renzini, A., Daddi, E., Dickinson, M., Cimatti, A., \& Tonini, C.\ 2010, \mnras, 407, 830 
\bibitem[Masjedi et al.(2006)]{Ma05} Masjedi, M., et al.\ 2006, \apj, 644, 54 
\bibitem[Masters, \etal (2003)]{M03} Masters, K. L., Giovanelli, R. \& Haynes, M. P. 2003, \aj, 126, 158. 
\bibitem[Masters et al. (2010a)]{M10dust} Masters, K. L., et al. 2010a, \mnras, 404, 792 
\bibitem[Masters et al. (2010b)]{M10} Masters, K. L., et al. 2010b, \mnras, 405, 783 
\bibitem[McCracken et al.(2007)]{2007ApJS..172..314M} McCracken, H.~J., et al.\ 2007, \apjs, 172, 314 
\bibitem[Naab et al.(2006)]{N06} Naab, T., Khochfar, S., \& Burkert, A.\ 2006, \apjl, 636, L81 
\bibitem[Padilla \& Strauss(2008)]{2008MNRAS.388.1321P} Padilla, N.~D., \& Strauss, M.~A.\ 2008, \mnras, 388, 1321 
\bibitem[Reyes et al.(2011)]{R11} Reyes, R., Mandelbaum,  R., Gunn, J.~E., Pizagno, J., \& Lackner, C.~N.\ 2011, arXiv:1106.1650 
\bibitem[Rix \& White(1989)]{RW89} Rix, H.-W.~R., \& White, S.~D.~M.\ 1989, \mnras, 240, 941 
\bibitem[Robaina et al.(2009)]{R09} Robaina, A.~R., et al.\ 2009, \apj, 704, 324 
\bibitem[Ross \& Brunner (2009)]{RB09}  Ross, A. J. \& Brunner, R. J. 2009, MNRAS, 399, 878
\bibitem[Ross et al.(2011)]{R11b} Ross, A.~J., et al. 2011b, \mnras ~(submitted; astroph/1105.2320). 
\bibitem[Sargent et al.(2007)]{sargent07} Sargent, M.~T., et al.\ 2007, \apjs, 172, 434  
\bibitem[Scarlata et al.(2007)]{ZEST} Scarlata, C., et al.\ 2007, \apjs, 172, 406 
\bibitem[Schawinski et al.(2007)]{2007MNRAS.382.1415S} Schawinski, K., Thomas, D., Sarzi, M., Maraston, C., Kaviraj, S., Joo, S.-J., Yi, S.~K., \& Silk, J.\ 2007, \mnras, 382, 1415 
\bibitem[Schawinski et al.(2009)]{S09} Schawinski, K., et  al.\ 2009, \mnras, 396, 818 
\bibitem[Schlegel et al.(1998)]{SFD98} Schlegel, D.~J., Finkbeiner, D.~P., \& Davis, M.\ 1998, \apj, 500, 525 
\bibitem[Scoville et al.(2007)]{Sc07} Scoville, N., et al.\ 2007, \apjs, 172, 38 
\bibitem[Shankar  \& Bernardi(2009)]{2009MNRAS.396L..76S} Shankar, F., \& Bernardi, M.\ 2009, \mnras, 396, L76 
\bibitem[Shankar et al.(2010)]{shankar2010} Shankar, F., Marulli, F., Bernardi, M., Boylan-Kolchin, M., Dai, X., \& Khochfar, S.\ 2010, \mnras, 405, 948 
\bibitem[Shankar et al.(2011)]{shankar2011} Shankar, F., Marulli, F., Bernardi, M., Mei, S., Meert, A., \& Vikram, V.\ 2011, arXiv:1105.6043 
\bibitem[Skibba et al.(2009)]{Sk09} Skibba, R.~A., et al.\ 2009, \mnras, 399, 966
\bibitem[Stoughton et al. (2002)]{S02} Stoughton, C., et al. 2002, AJ, 123, 485 
\bibitem[Strateva et al.(2001)]{2001AJ....122.1861S} Strateva, I., et al.\ 2001, \aj, 122, 1861 
\bibitem[Thomas et al.(2010)]{2010MNRAS.404.1775T} Thomas, D., Maraston, 
C., Schawinski, K., Sarzi, M., \& Silk, J.\ 2010, \mnras, 404, 1775
\bibitem[Tojeiro \& Percival(2010)]{2010MNRAS.405.2534T} Tojeiro, R., \& Percival, W.~J.\ 2010, \mnras, 405, 2534 
\bibitem[Trujillo, Graham \& Caon (2001)]{TGC01} Trujillo, I., Graham, A. \& Caon, N. 2001, \mnras, 326, 869
\bibitem[Trujillo et al. (2006)]{trujillo06} Trujillo I. et al., 2006a, MNRAS, 373, L36 
\bibitem[Wake et al.(2008)]{2008MNRAS.387.1045W} Wake, D.~A., et al.\ 2008, \mnras, 387, 1045 
\bibitem[White et al. (2011)]{White11} White, M. et al. 2011, ApJ, 728, 126. 
\bibitem[Wolf et al.(2009)]{W09} Wolf, C., et al.\ 2009, \mnras, 393, 1302 
\bibitem[van der Wel et al.(2009)]{2009ApJ...698.1232V} van der Wel, A., Bell, E.~F., van den Bosch, F.~C., Gallazzi, A., \& Rix, H.-W.\ 2009, \apj, 698, 1232 
\bibitem[van der Wel et a. (2008)]{vanderwel2008} van derWel A., Holden B. P., Zirm A.W., FranxM., Rettura A., Illingworth G. D., Ford H. C., 2008, ApJ, 688, 48 
\bibitem[van Dokkum et al.(1999)]{vd99} van Dokkum, P.~G., Franx, M., Fabricant, D., Kelson, D.~D., \& Illingworth, G.~D.\ 1999, \apjl, 520, L95 
\bibitem[van Dokkum et al.(2008)]{vandokkum08} van Dokkum P. G. et al., 2008, ApJ, 677, L5 
\bibitem[Yip et al.(2011)]{Y11} Yip, C.-W., Szalay, A.~S., Carliles, S., \& Budavari, T.\ 2011, ApJ, 730, 54
\bibitem[York et al.(2000)]{Y00} York, D.~G., et al.\ 2000, \aj, 120, 1579 
\bibitem[Zehavi et al. (2005)]{Z05} Zehavi, I et al. 2005, ApJ 630, 1
\end{thebibliography}
\end{document}